\documentclass[conference]{IEEEtran}
\IEEEoverridecommandlockouts
\usepackage{cite}
\usepackage{amsmath,amssymb,amsfonts}
\usepackage{algorithmic}
\usepackage{graphicx}
\usepackage{textcomp}
\usepackage{xcolor}
\usepackage{hyperref}
\usepackage{tabularx}
\usepackage{pifont}
\usepackage{makecell}
\usepackage{multirow}
\usepackage{booktabs}
\usepackage{bbm}
\usepackage{threeparttable}

\newcommand{\cmark}{\ding{51}}
\newcommand{\xmark}{\ding{55}}

\def\BibTeX{{\rm B\kern-.05em{\sc i\kern-.025em b}\kern-.08em
    T\kern-.1667em\lower.7ex\hbox{E}\kern-.125emX}}
\begin{document}


\title{LymphNode: A Plug-and-Play Access Control Method for Deep Neural Networks%
\thanks{\copyright~2026 IEEE. Personal use of this material is permitted. 
Permission from IEEE must be obtained for all other uses, in any current or 
future media, including reprinting/republishing this material for advertising 
or promotional purposes, creating new collective works, for resale or 
redistribution to servers or lists, or reuse of any copyrighted component of 
this work in other works.}
}

\author{
\IEEEauthorblockN{Hanyu Pei}
\IEEEauthorblockA{Department of Computer Science\\
and Engineering\\
University of Louisville\\
hanyu.pei@louisville.edu}
\and
\IEEEauthorblockN{Shang Liu}
\IEEEauthorblockA{Department of Computer Science\\
and Engineering\\
University of Louisville\\
shang.liu@louisville.edu}
\and
\IEEEauthorblockN{Zeyan Liu}
\IEEEauthorblockA{Department of Computer Science\\
and Engineering\\
University of Louisville\\
zeyan.liu@louisville.edu}
}

\maketitle

\begin{abstract}
Deep Neural Networks (DNNs) are high-value intellectual property (IP), yet deploying them to edge environments exposes them to \textbf{unrestricted oracle access}, rendering them vulnerable to model extraction and inversion attacks. Existing defenses fail to address this practically: passive watermarking only offers post-hoc provenance, while active defenses impose prohibitive latency or require persistent access to sensitive training data. To bridge this gap, we propose \textit{LymphNode}, a novel post-hoc defense framework that acts as an intrinsic ``immune system" within the model. \textit{LymphNode} enforces a strict ``default-deny'' policy: it actively neutralizes model utility for unauthorized queries via \textbf{Generalized Sparse Universal Adversarial Perturbations (GSUAP)} injected into the feature space, effectively blocking gradient estimation and data inference. Utility is selectively restored only for authorized inputs carrying a stealthy feature-domain credential. Our framework is highly practical: it is \textbf{data-efficient}, establishing robust protection with fewer than 100 samples ($<1\%$ of training data), and \textbf{cross-dataset adaptable}, enabling protection using public surrogate datasets. \textit{LymphNode} thus provides a lightweight, immediately deployable defense for high-stakes scenarios where original training data is restricted or unavailable.
\end{abstract}

\begin{IEEEkeywords}
Model IP Protection, Active Defense, Model Extraction.
\end{IEEEkeywords}

\section{Introduction}

Training modern machine learning models is an exceptionally resource-intensive endeavor. Large-scale foundation models, such as GPT-4 or LLaMA~\cite{openai2024gpt4,touvron2023llama}, require months of computation, making their weights invaluable intellectual property (IP). However, to satisfy latency and privacy requirements, models are frequently deployed to edge devices or on-premise servers. While this decentralization protects raw data, it inadvertently grants adversaries unrestricted oracle access to the model interface. Unlike rate-limited cloud APIs, edge deployment allows attackers to query the model with infinite volume at zero latency. Exploiting this, adversaries can launch sophisticated model extraction attacks~\cite{tramer2016stealing, jagielski2020high} to reverse-engineer parameters or model inversion attacks~\cite{fredrikson2015model} to infer sensitive training data. This capability allows malicious actors to functionally replicate proprietary models without needing to decrypt the underlying static files. Despite these risks, on-premise deployment remains non-negotiable for sensitive industries (e.g., healthcare, finance), creating a direct conflict between operational necessity and IP security.

Existing defense mechanisms struggle to resolve this conflict due to significant practicality limitations. \textit{Passive defenses}, such as watermarking~\cite{adi2018turning,uchida2017embedding,zhang2018protecting}, primarily serve copyright provenance \textit{after} theft occurs, failing to proactively prevent functional extraction. Conversely, \textit{active defenses} aim to enforce access control but face critical bottlenecks. Cryptographic methods (e.g., Deep-Lock~\cite{alam2020deeplock}) incur prohibitive computational overhead, rendering them unsuitable for real-time edge applications. Meanwhile, structural authorization methods~\cite{chen2021model,xue2024ssat,xue2021advparams} embed locking mechanisms but suffer from severe data dependency: they typically require computationally expensive full model retraining or persistent access to the complete original dataset. This makes them infeasible for real-world scenarios where training data is sensitive, legally restricted (e.g., GDPR), or unavailable post-training. More recently, output perturbation methods~\cite{orekondy2020prediction,zhang2023cip} actively degrade extracted model quality at the prediction interface, yet they require a runtime interposition layer inherent to cloud-API deployments and cannot protect models physically transferred to edge environments.

In this paper, we propose \textit{LymphNode}, a post-hoc plugin framework designed to bridge these practicality gaps. Unlike invasive methods requiring retraining, our approach integrates a lightweight ``checkpoint" directly into the inference pipeline. Drawing inspiration from the biological immune system, \textit{LymphNode} enforces a rigorous ``default-deny" policy. It treats all inputs as unauthorized and injects a pre-computed, Generalized Sparse Universal Adversarial Perturbation (GSUAP, detailed in Sec.~\ref{sec: GSUAP}) into critical channels. This defense persistently neutralizes the model's output quality for arbitrary queries, thereby significantly degrading the quality of outputs available for model extraction~\cite{tramer2016stealing,orekondy19knockoff} and inversion~\cite{fredrikson2015model,zhang2020secret} attacks. Access is restored only when an authorized input with a specific feature-domain credential is verified, triggering an inverse perturbation to cancel the noise. Consequently, the model remains useless to adversaries relying on oracle access, while authorized users transparently recover full fidelity. The overall architecture is illustrated in Fig~\ref{fig:framework}. In summary, our key contributions are:

\begin{figure*}[!t]  
    \centering
    \includegraphics[width=0.99\textwidth]{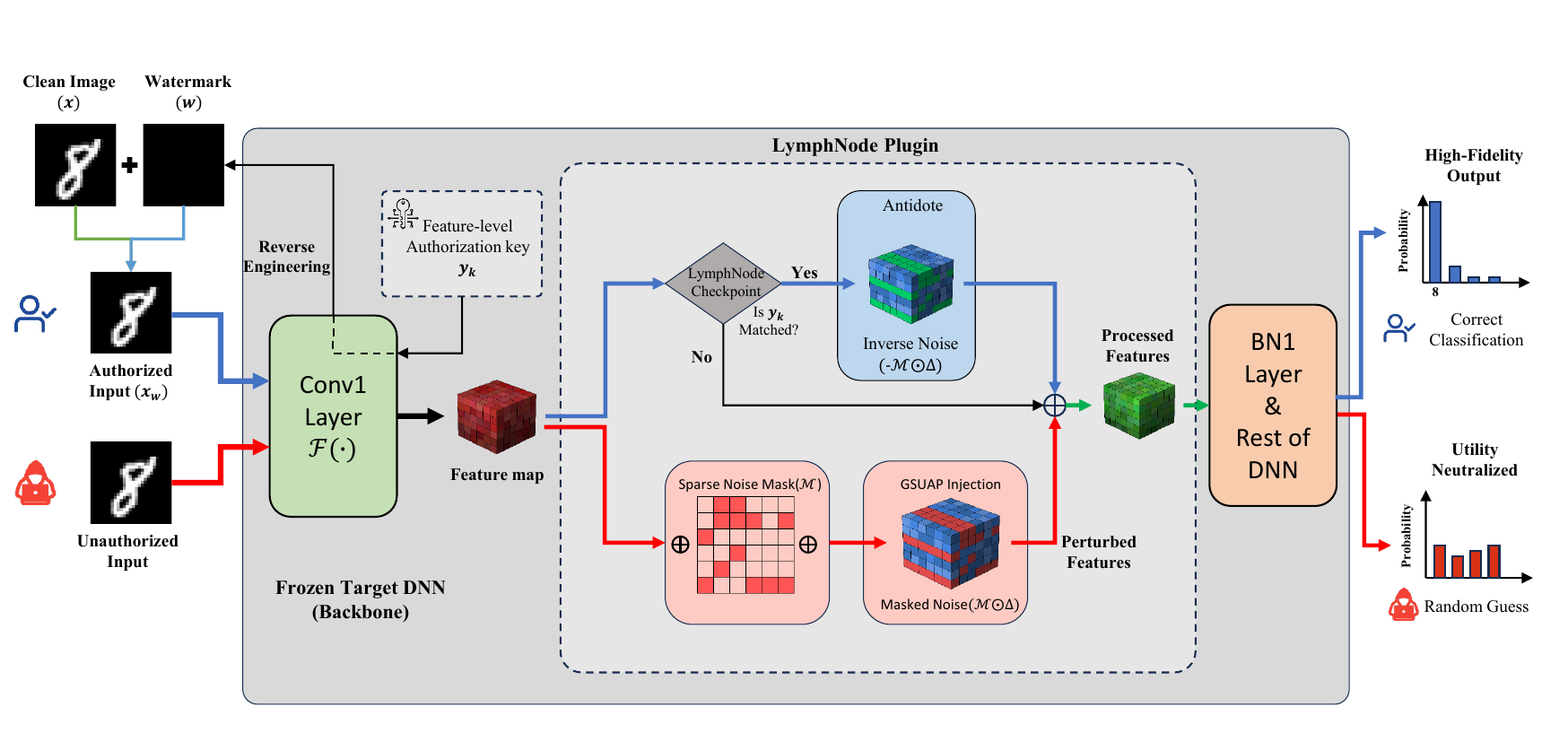}
    \caption{An overview of LymphNode plugin}
    \label{fig:framework}
\end{figure*}

\begin{itemize} 
    \item We propose \textit{LymphNode}, a novel post-hoc plugin framework that provides active IP protection. It actively neutralizes models against unauthorized input via a sparse adversarial perturbation, while restoring full fidelity for authorized users through a stealthy, feature-domain verification protocol.

    \item Our framework demonstrates exceptional \textbf{data adaptivity and efficiency}. We prove that the framework significantly lowers the deployment barrier: it can be robustly initialized using as few as 50-100 samples ($<1\%$ of the training set) and is capable of protecting target models using public surrogate datasets (e.g., protecting STL-10 models with CIFAR-10 noise). This eliminates the strict dependency on original private data required by prior art.

    \item We conduct a comprehensive security and robustness evaluation. We rigorously validate the system's resilience against adaptive threats, including deep generative credential forgery and model hijacking via fine-tuning. Furthermore, we demonstrate that the protection mechanism remains dependable under real-world distortions (e.g., lossy compression), while maintaining near-perfect watermark imperceptibility (LPIPS $\approx 0.001$), confirming its suitability for practical deployment.
\end{itemize}

The remainder of this paper is organized as follows. Section~\ref{sec:related_work} reviews related work in model watermarking and active defenses. Section~\ref{sec:Method} details the architecture and the post-hoc training mechanism of \textit{LymphNode}. Section~\ref{sec:Experiment} presents the experimental setup and a comprehensive evaluation of effectiveness, efficiency, and robustness. We further discuss the data adaptivity and robustness of our framework in Sec.~\ref{sec:data adaptivity} and Sec.~\ref{sec:robustness}. Finally, we conclude the paper in Section~\ref{sec:conclusion}.

\section{Related Work}
\label{sec:related_work}

\subsection{Passive Defense: IP Provenance}
Passive defense mechanisms for DNN intellectual property protection predominantly consist of model watermarking and fingerprinting techniques. Existing watermarking approaches~\cite{uchida2017embedding,adi2018turning,rouhani2019deepsigns} embed identifying information directly into model parameters~\cite{uchida2017embedding}, activation maps~\cite{rouhani2019deepsigns}, or couple them with backdoor triggers~\cite{adi2018turning,li2022untargeted,guo2023domain} to enable ownership verification. In parallel, fingerprinting methods~\cite{cao2021ipguard,lemerrer2020adversarial} extract intrinsic model attributes, such as decision boundary characteristics or specific prediction behaviors on adversarial examples, to construct unique identifiers without altering the original model weights. Collectively, these embedded or extracted signatures allow model owners to verify authorship through statistical analysis or behavioral querying. More recently, DynaMarks~\cite{chakraborty2022dynamarks} embeds transferable watermarks into extracted surrogates by dynamically altering output probabilities at inference time, yet the extracted model retains full functionality and the watermark serves only for post-hoc ownership claims.

However, these approaches fundamentally constitute a passive line of defense. Their primary utility is restricted to copyright provenance which serves to prove ownership after a theft has already occurred, typically for litigation purposes. As comprehensively analyzed by Lukas et al.~\cite{lukas2022sok}, while such methods can verify ownership with high confidence, they possess no capability to proactively prevent unauthorized model execution. Consequently, once a protected model is leaked or distributed, adversaries can freely deploy and monetize the asset, leaving the window of exploitation entirely open despite the presence of ownership proofs.

\subsection{Recent Advances in UAP}
Recent scholarship in adversarial learning has produced highly effective strategies for generating universal adversarial perturbations (UAP). Techniques such as SGA~\cite{liu2023enhancing}, DM-UAP~\cite{zhang2025improving}, and RobustUAP~\cite{xu2022robust} introduce sophisticated optimization objectives, including stochastic gradient aggregation and dynamic maximin frameworks, to maximize the transferability and robustness of attacks across diverse models and distributions. Parallel research like Sparse-PGD~\cite{zhong2024sparse} has further refined the generation of sparse adversarial noise. 

However, these methods are fundamentally engineered for offensive generalization rather than defensive controllability. Their primary goal is to construct an irreversible perturbation that degrades performance across unknown target domains, a process that typically necessitates extensive training data to approximate universal vulnerability manifolds. In the context of active IP protection via a lightweight plugin framework, the operational objective is fundamentally different. We require a targeted neutralization mechanism capable of locking a specific, known model with minimal setup costs. Consequently, our work adopts the Generalized Data-Free UAP (GD-UAP)~\cite{mopuri2018generalizable} formulation. By leveraging its strong data-independent characteristics, we can efficiently optimize the perturbation to saturate the target model's feature space without the prohibitive data dependency associated with transferability-oriented attacks.

\begin{table}[!t]
    \centering
    \caption{Qualitative comparison with state-of-the-art active IP protection methods. \textit{LymphNode} uniquely combines data independence, zero-retraining, and negligible overhead.}
    \label{tab:sota_comparison}
    \small
    \setlength{\tabcolsep}{4pt} 
    \renewcommand{\arraystretch}{1.2}
    \begin{tabular}{lcccc}
        \toprule
        \textbf{Method} & \textbf{Origin Data} & \textbf{Retrain} & \textbf{Post-Hoc} & \textbf{Overhead} \\
        \midrule
        Deep-Lock~\cite{alam2020deeplock}    & \xmark & \xmark & \cmark & High   \\
        AdvParams~\cite{xue2021advparams}    & \cmark & \cmark & \xmark & Low    \\
        SSAT~\cite{xue2024ssat}              & \cmark & \cmark & \xmark & Low    \\
        IDEA~\cite{chakraborty2020hpnn}      & \cmark & \cmark & \xmark & Medium \\
        ModelLock~\cite{gong2024modellock}   & \cmark & \cmark & \xmark & Low    \\
        \midrule
        \textbf{Ours} & \textbf{\xmark} & \textbf{\xmark} & \textbf{\cmark} & \textbf{Negligible} \\
        \bottomrule
    \end{tabular}
\end{table}

\subsection{Active Defense: Cryptographic Approaches}
An alternative paradigm involves cryptographic techniques that provide active protection during inference. Hardware-based solutions like Trusted Execution Environments (TEEs)~\cite{tramer2019slalom} and algorithmic solutions like Homomorphic Encryption (HE)~\cite{natarajan2023chexmix} theoretically guarantee secure execution. Similarly, methods like Deep-Lock~\cite{alam2020deeplock} and NN-Lock~\cite{alam2022nnlock} propose S-Box-based parameter encryption, requiring decryption for every query. However, recent surveys~\cite{feng2024survey} highlight significant practical barriers: TEEs are susceptible to side-channel attacks, while HE and parameter decryption schemes impose prohibitive computational overhead and latency penalties. This renders them impractical for resource-constrained edge deployments where real-time inference is non-negotiable.

\subsection{Active Defense: Structural Authorization}
More recent works aim to achieve active authorization control by modifying the model structure or weights. Representative methods include ModelGuard~\cite{chen2021model}, ActiveGuard~\cite{xue2023activeguard}, SSAT~\cite{xue2024ssat}, and AdvParams~\cite{xue2021advparams}. These approaches embed backdoors or adversarial perturbations into the weights during training to discriminate between authorized and unauthorized users. Recent advancements like ModelLock~\cite{gong2024modellock} leverage diffusion models to edit the training distribution for enhanced locking.

Crucially, these approaches suffer from a severe \textbf{setup cost bottleneck}. They typically require either: (i) computationally expensive full model retraining (e.g., ModelLock, SSAT), or (ii) persistent access to the complete original training dataset (e.g., AdvParams, IDEA~\cite{chakraborty2020hpnn}) to compute gradients or generate triggers. This requirement is often infeasible in real-world scenarios due to strict privacy regulations (e.g., GDPR) or data loss. Even sample-specific approaches remain bound to the original training pipeline. To clearly distinguish our contribution within the active defense landscape, we provide a qualitative comparison in Table~\ref{tab:sota_comparison}. 

\subsection{Active Defense: Anti-Extraction via Output Perturbation}
\label{sec:related_anti_extraction}

A parallel line of work actively degrades extracted models by perturbing responses at the prediction interface. Methods like Prediction Poisoning~\cite{orekondy2020prediction}, CIP~\cite{zhang2023cip}, and AMAO~\cite{jiang2024amao} dynamically inspect, score, or poison per-query outputs to disrupt attacker objectives.

These methods strictly require a server-side interposition layer---a capability inherent to cloud-APIs but physically infeasible at the edge, where adversaries directly invoke the forward pass. In contrast, \textit{LymphNode} embeds protection statically within the computational graph as a feature-space intervention, eliminating runtime query analysis.

\section{Framework}
\label{sec:Method}

In this section, we present \textit{LymphNode}, a post-hoc defense framework that structurally integrates active IP protection directly into the target Deep Neural Network (DNN). Unlike external pre-processing wrappers, our approach fuses the security mechanism into the model's computational graph as an intrinsic intervention node, specifically targeting the intermediate feature space. We operate within a standard black-box deployment scenario where the target model is encapsulated in an inference environment that preserves runtime integrity. The framework governs the inference logic by enforcing a strict ``default-deny'' policy embedded within the forward pass topology: a \textit{Generalized Sparse Universal Adversarial Perturbation} (GSUAP, detailed in Sec.~\ref{sec: GSUAP}) is injected into the latent features by default, neutralizing model utility for any unauthorized access. Full fidelity is only recovered when a coupled antidote mechanism validates a stealthy feature-domain credential, thereby rendering the security logic inseparable from the model's fundamental feature extraction process.

\subsection{Threat Model and Assumptions}
\label{sec:threat_model}

A three-party edge deployment scenario is considered. The \textit{Model Owner} trains the target DNN, optimizes the \textit{LymphNode} GSUAP plugin, and issues authorization keys to the \textit{Edge Operator} via a secure out-of-band channel. The \textit{Edge Operator} deploys the protected model within a trusted runtime environment (e.g., signed firmware) that preserves parameter integrity but exposes model functionality through an inference API. Crucially, the \textit{Edge Operator} independently manages key distribution to \textit{Authorized End-Users} within their trust domain.

The adversary model is strictly gray-box: the adversary possesses knowledge of the model architecture and the general defense mechanism, but lacks access to (i)~the secret authorization key $\mathbf{k}$, (ii)~runtime memory or intermediate feature maps, and (iii)~gradient or backward-pass computations. Oracle access is therefore the only available attack surface: the adversary may submit arbitrary queries---including adaptive queries crafted from previous outputs---and observe the returned probability vectors.

\subsection{Authorization via Feature-Domain Verification}
\label{sec:identify}

As the core decision logic of the \textit{LymphNode} framework (Fig.~\ref{fig:framework}), this module implements identity verification by embedding a discrete credential into the continuous feature space. The authorization key is defined as a secret $N$-bit binary string $\mathbf{k} \in \{0,1\}^N$. To embed this key, we select $N$ carrier features from the first convolutional layer's output using a distributed strategy and denote the feature representation of authorization key as $\mathbf{y}_k$. Assuming the layer comprises $r$ kernels with a spatial size of $m \times n$, we select $v$ distinct kernels and $h$ distinct spatial locations within each kernel, such that the total capacity is $N = v \times h$ (noticing that these $h$ features from the same feature map correspond to non-overlapping pixel-domain regions). Based on the ablation study in prior work~\cite{liu2022loneneuron}, we adopt a configuration of $N = 32$ and $v = 4$ to balance capacity and stealthiness. 

For each carrier feature $y$, we target the $s$-th bit after the binary point for verification. Mathematically, the extraction of the verification bit $b$ is formulated as:
\begin{equation}
b = \lfloor |y| \cdot 2^s \rfloor \bmod 2
\end{equation}
Intuitively, a larger $s$ implies that the verification relies on finer-grained quantization noise, resulting in smaller perturbations and better stealthiness. Following Liu et al.~\cite{liu2022loneneuron}, we set $s = 6$, rendering the modifications visually imperceptible in the pixel domain.

To pass this verification, an authorized user must generate an input $\mathbf{x}_{auth} = \mathbf{x} + \mathbf{w}$ such that its feature representation matches $\mathbf{k}$. This constitutes an inverse problem: finding a perturbation $\mathbf{w}$ that satisfies:
\begin{equation}
\label{eq:inverse}
    \mathcal{F}(\mathbf{x} + \mathbf{w}) = \mathbf{y}_{\mathbf{k}}
\end{equation}
where $\mathcal{F}$ represents the first convolutional layer. Since direct analytical inversion is ill-posed due to dimensionality mismatch, we employ a search-based strategy. We first locate the receptive fields corresponding to the $N$ features and then employ a random search to identify a valid $\mathbf{w}$. The computational complexity is $O(h \times 2^v)$, which is highly efficient given the bound $N = v \times h$. Empirical results in Sec.~\ref{sec:overhead} confirm that generating 1,000 unique credentials requires less than 2 seconds.

Finally, the verification mechanism necessitates a fundamental security trade-off. The choice of $N$ balances efficiency and specificity. A smaller $N$ accelerates generation but increases the risk of an unauthorized input coincidentally matching $\mathbf{k}$ (a ``collision"), which would grant illicit access. The theoretical collision probability $P_c$ is estimated as follows:
\begin{equation}
\label{eq:collision}
    P_c = P(b_1=k_1, \dots, b_N=k_N) = \prod_{i=1}^{N} P(b_i=k_i) = 2^{-N}
\end{equation}
where $P(b_i = k_i)$ denotes the collision probability for the $i^{th}$ bit. The Eq.~\ref{eq:collision} holds when $P(k_i = 0) = P(k_i = 1) = 0.5$ and $P(k_i) = P(k_i|k_j), \forall i \neq j$, indicating the bit-wise collision follows binomial distribution. These conditions are verified to hold by experiments in ~\cite{liu2022loneneuron}. For our setting ($N=32$), $P_c \approx 2.33 \times 10^{-10}$. This infinitesimal probability ensures that the ``default-deny" policy is robustly enforced, preventing accidental authorization by benign inputs. 

\subsection{Model Performance Neutralization via GSUAP}
\label{sec: GSUAP}

To efficiently regulate the model's fidelity for unauthorized access, we propose \textbf{Generalized Sparse Universal Adversarial Perturbations (GSUAP)}. While standard Universal Adversarial Perturbations (UAPs)~\cite{moosavi2017universal, mopuri2018generalizable} seek to fool a model on all inputs, applying them indiscriminately increases computational overhead and detection risk. Therefore, our objective is to adapt structured pruning principles~\cite{molchanov2016pruning, li2016pruning} to identify a minimal subset of \textit{decision-critical channels}, where targeted noise injection can maximally disrupt model behavior while minimizing the modification footprint.

Performance Neutralization operates in two sequential phases: first, channels are selected based on their gradient sensitivity regarding the classification loss; second, a constant adversarial perturbation is optimized on these selected channels.

\textbf{Phase 1: Weight Gradient-based Channel Selection.} 
Given a pre-trained clean model $f_\theta$ and a small calibration dataset $\mathcal{D}_{cal}$, we quantify each channel's importance. Drawing from gradient-based pruning criteria~\cite{molchanov2016pruning}, we define the importance score for the $j$-th channel in a layer (with weight tensor $W$) as the expected gradient magnitude:
\begin{equation}
\text{Score}_j = \mathbb{E}_{(x,y) \sim \mathcal{D}_{cal}} \left[ \left\| \frac{\partial \mathcal{L}_{CE}(f_\theta(x), y)}{\partial W_j} \right\|_2 \right]
\end{equation}
where $W_j$ denotes the kernel weights for channel $j$. A high score indicates that the loss is highly sensitive to variations in this channel. Given a target sparsity ratio $r \in (0, 1]$, we select the top-$k$ ($k = \lfloor r \cdot M \rfloor$) channels to construct a binary mask $\mathcal{M} \in \{0,1\}^M$. Unlike magnitude-based ranking~\cite{han2015learning} which evaluates static weights, this gradient-based criterion dynamically identifies the optimal ``acupuncture points" for performance neutralization.

\textbf{Phase 2: Sparse Adversarial Noise Optimization.} 
With the channel mask $\mathcal{M}$ fixed and model parameters frozen, we optimize a universal additive noise $\Delta$ to maximize misclassification on unauthorized inputs. For any input $x$, the noise is injected into the selected channels of the feature map:
\begin{equation}
    \tilde{F}(x) = F(x) + \mathcal{M} \odot \Delta
\end{equation}
where $F(x)$ is the clean feature map and $\odot$ denotes element-wise multiplication. To find the optimal perturbation, we employ Projected Gradient Ascent (PGA) to maximize the classification loss $\mathcal{L}_{CE}$. Let $g^{(t)}$ denote the gradient of the loss with respect to the noise at step $t$, i.e., $g^{(t)} = \nabla_{\Delta} \mathcal{L}_{CE}(f(x; \Delta^{(t)}), y)$. The update rule is formulated as:
\begin{equation}
    \Delta^{(t+1)} = \Pi_{\epsilon} \left( (\Delta^{(t)} + \alpha \cdot \text{sign}(g^{(t)})) \odot \mathcal{M} \right)
\end{equation}
where $\alpha$ is the step size, $\Pi_{\epsilon}$ projects the noise onto the $\ell_\infty$-ball bounded by $\epsilon$. Crucially, this optimization targets the unauthorized scenario: the goal is to find a single, static perturbation $\Delta$ that, when masked by $\mathcal{M}$, causes the model to fail on clean inputs. Once optimized, this noise module is integrated into the \textit{LymphNode} plugin as shown in Fig.~\ref{fig:framework}. It remains active by default to neutralize performance for unauthorized users, and is only bypassed (via the antidote mechanism, an inverse GSUAP) when a valid feature-domain credential is verified.


\section{Experimentation}
\label{sec:Experiment}

In this section, we establish the foundational performance of \textit{LymphNode}, focusing on its core capability to regulate model inference under realistic deployment constraints. We rigorously evaluate the framework along three primary dimensions, beginning with the neutralization effectiveness, where we verify the system's ability to selectively suppress unauthorized accuracy across diverse architectures and datasets. Concurrently, we assess the neutralization efficiency to quantify the protection gain relative to the structural modification cost. To ensure practical viability, we also analyze system performance by benchmarking computational overheads, including latency, throughput, and memory, validating the framework's suitability for resource-constrained environments. Comprehensive analyses regarding design choices, data dependencies, and security resilience are presented in subsequent sections.

\subsection{Neutralization Effectiveness Evaluation}
\label{sec:exp_effectiveness}

\begin{table*}[!t]
    \centering
    \tiny 
    \caption{Neutralization effectiveness across datasets and architectures.\textsuperscript{$\dagger$}}
    \label{tab:full_comparison_gap_uniform}
    \renewcommand{\arraystretch}{0.9}
    \resizebox{\textwidth}{!}{
\begin{tabular}{l c | cccc | cccc | cccc}
            \toprule[1pt]
            \multirow{2}{*}{\textbf{Model}} & \multirow{2}{*}{\textbf{Ratio (\%)}} & 
            \multicolumn{4}{c|}{\textbf{CIFAR-10}} & 
            \multicolumn{4}{c|}{\textbf{MNIST}} & 
            \multicolumn{4}{c}{\textbf{SVHN}} \\
            \cline{3-14} 
            \rule{0pt}{2.2ex}
             & & Gauss & SUAP & GSUAP & VIP & Gauss & SUAP & GSUAP & VIP & Gauss & SUAP & GSUAP & VIP \\
            \midrule
            
            \multirow{5}{*}{\shortstack{ResNet-18 }} 
             & 20  & 93.6 & \underline{93.2} & \textbf{88.6} & 94.5 & 49.0 & \textbf{10.2} & \textbf{10.2} & 99.6 & 64.6 & \underline{13.2} & \textbf{8.1} & 96.1 \\
             & 40  & 87.4 & \underline{81.4} & \textbf{36.4} & 94.5 & 31.3 & \underline{11.3} & \textbf{11.3} & 99.6 & 60.3 & \underline{11.1} & \textbf{7.4} & 96.1 \\
             & 60  & 85.4 & \underline{72.0} & \textbf{13.6} & 94.5 & 29.4 & \textbf{9.3}  & \textbf{9.3}  & 99.6 & 50.8 & \underline{10.9} & \textbf{7.2} & 96.1 \\
             & 80  & 80.2 & \underline{67.0} & \textbf{11.0} & 94.5 & 24.7 & \underline{10.6} & \textbf{10.2} & 99.6 & 46.5 & \underline{10.8} & \textbf{6.6} & 96.1 \\
             & 100 & 73.6 & \underline{48.4} & \textbf{10.6} & 94.5 & 22.6 & \textbf{10.2} & \textbf{10.2} & 99.6 & 44.0 & \underline{10.8} & \textbf{6.6} & 96.1 \\
            \midrule
            
            \multirow{5}{*}{\shortstack{ResNet-50 }} 
             & 20  & 95.2 & \underline{87.2} & \textbf{79.2} & 95.8 & 92.6 & \underline{16.7} & \textbf{16.3} & 99.8 & 68.4 & \underline{25.3} & \textbf{10.8} & 96.6 \\
             & 40  & 83.8 & \underline{65.2} & \textbf{47.8} & 95.8 & 45.7 & \underline{11.1} & \textbf{10.9} & 99.8 & 65.8 & \underline{21.4} & \textbf{9.5}  & 96.6 \\
             & 60  & 74.2 & \underline{47.8} & \textbf{25.6} & 95.8 & 35.3 & \underline{10.4} & \textbf{10.2} & 99.8 & 59.4 & \underline{19.4} & \textbf{9.8}  & 96.6 \\
             & 80  & 72.6 & \underline{42.0} & \textbf{11.2} & 95.8 & 33.7 & \underline{10.4} & \textbf{10.2} & 99.8 & 52.7 & \underline{19.2} & \textbf{9.5}  & 96.6 \\
             & 100 & 68.6 & \underline{32.4} & \textbf{9.0}  & 95.8 & 32.0 & \underline{10.4} & \textbf{10.0} & 99.8 & 48.6 & \underline{19.2} & \textbf{9.4}  & 96.6 \\
            \midrule
            
            \multirow{5}{*}{\shortstack{ViT-Tiny }} 
             & 20  & 71.6 & \underline{49.0} & \textbf{27.8} & 91.8 & 46.3 & \underline{11.2} & \textbf{10.4} & 99.3 & 67.8 & \textbf{19.8} & \underline{24.1} & 96.3 \\
             & 40  & 51.0 & \underline{16.6} & \textbf{11.6} & 91.8 & 32.4 & \underline{10.9} & \textbf{10.0} & 99.3 & 43.0 & \underline{18.4} & \textbf{18.1} & 96.3 \\
             & 60  & 39.8 & \underline{13.6} & \textbf{11.8} & 91.8 & 23.8 & \underline{9.4}  & \textbf{8.9}  & 99.3 & 32.5 & \underline{11.2} & \textbf{10.8} & 96.3 \\
             & 80  & 34.8 & \underline{11.8} & \textbf{10.6} & 91.8 & 21.2 & \underline{9.9}  & \textbf{8.4}  & 99.3 & 26.0 & \textbf{10.3} & \underline{10.8} & 96.3 \\
             & 100 & 29.8 & \underline{11.2} & \textbf{10.6} & 91.8 & 19.0 & \underline{9.8}  & \textbf{8.4}  & 99.3 & 22.2 & \textbf{10.2} & \underline{10.8} & 96.3 \\
            \midrule
            
            \multirow{5}{*}{\shortstack{ViT-Small }} 
             & 20  & 55.6 & \underline{22.6} & \textbf{13.6} & 89.4 & 40.8 & \underline{10.8} & \textbf{10.7} & 99.2 & 43.7 & \underline{23.6} & \textbf{21.8} & 97.6 \\
             & 40  & 36.0 & \underline{15.2} & \textbf{10.2} & 89.4 & 26.1 & \textbf{10.0} & \textbf{10.0} & 99.2 & 25.9 & \underline{22.0} & \textbf{21.0} & 97.6 \\
             & 60  & 27.0 & \underline{11.2} & \textbf{10.0} & 89.4 & 22.7 & \underline{9.8}  & \textbf{9.4}  & 99.2 & 20.0 & \textbf{18.2} & \underline{18.9} & 97.6 \\
             & 80  & 24.8 & \underline{10.6} & \textbf{10.0} & 89.4 & 17.9 & \textbf{9.5}  & \textbf{9.5}  & 99.2 & 20.0 & \textbf{13.2} & \underline{14.6} & 97.6 \\
             & 100 & 23.6 & \underline{11.6} & \textbf{10.0} & 89.4 & 14.1 & \underline{9.5}  & \textbf{9.4}  & 99.2 & 20.0 & \underline{13.2} & \textbf{11.0} & 97.6 \\
            \midrule
            
            \multirow{5}{*}{\shortstack{DenseNet }} 
             & 20  & 14.2 & \underline{13.5} & \textbf{11.3} & 95.8 & 12.3 & \underline{10.7} & \textbf{9.5} & 99.5 & 15.8 & \underline{15.6} & \textbf{9.9}  & 96.1 \\
             & 40  & 10.7 & \underline{10.7} & \textbf{9.6}  & 95.8 & 10.1 & \underline{9.8}  & \textbf{9.8} & 99.5 & 15.4 & \underline{14.6} & \textbf{8.0}  & 96.1 \\
             & 60  & 10.6 & \underline{9.6}  & \textbf{9.3}  & 95.8 & 10.3 & \underline{10.3} & \textbf{9.3} & 99.5 & 13.1 & \underline{14.1} & \textbf{7.7}  & 96.1 \\
             & 80  & 10.2 & \underline{9.7}  & \textbf{9.0}  & 95.8 & 10.5 & \textbf{9.5}  & \textbf{9.5} & 99.5 & 10.5 & \underline{9.4}  & \textbf{7.6}  & 96.1 \\
             & 100 & 10.2 & \underline{9.5}  & \textbf{9.0}  & 95.8 & 10.4 & \underline{9.4}  & \textbf{9.4} & 99.5 & 9.9  & \underline{8.5}  & \textbf{7.7}  & 96.1 \\
            \midrule

            \multirow{5}{*}{\shortstack{AlexNet }} 
             & 20  & 23.5 & \underline{12.7} & \textbf{11.0} & 90.5 & 12.6 & \underline{9.8}  & \textbf{9.7}  & 98.9 & 28.0 & \underline{15.7} & \textbf{10.5} & 95.4 \\
             & 40  & 20.8 & \underline{11.9} & \textbf{11.5} & 90.5 & 11.5 & \textbf{9.4}  & \underline{9.5}  & 98.9 & 27.3 & \underline{11.7} & \textbf{7.3}  & 95.4 \\
             & 60  & 12.7 & \underline{9.9}  & \textbf{9.8}  & 90.5 & 10.9 & \underline{9.2}  & \textbf{9.0}  & 98.9 & 26.8 & \underline{8.0}  & \textbf{6.0}  & 95.4 \\
             & 80  & 11.0 & \underline{9.2}  & \textbf{9.0}  & 90.5 & 10.6 & \underline{9.2}  & \textbf{8.9}  & 98.9 & 20.7 & \underline{7.1}  & \textbf{6.0}  & 95.4 \\
             & 100 & 11.4 & \underline{9.2}  & \textbf{9.0}  & 90.5 & 10.6 & \underline{9.2}  & \textbf{8.9}  & 98.9 & 16.7 & \underline{7.1}  & \textbf{6.0}  & 95.4 \\
             
            \bottomrule[1pt]
        \end{tabular}}
    \vspace{2pt}
    \parbox{\textwidth}{\footnotesize \textsuperscript{$\dagger$}\textbf{Ratio}: percentage of channels with GSUAP injection. Gauss/SUAP/GSUAP: accuracy of unauthorized inputs (\%, lower is better). VIP: accuracy of authorized inputs (\%, higher is better). \textbf{Bold}: best suppression; \underline{underline}: second best.}
\end{table*}

In this section, we evaluate the capability of \textit{LymphNode} to selectively neutralize model performance for unauthorized access. As detailed in Sec.~\ref{sec: GSUAP}, our framework injects Generalized Sparse Universal Adversarial Perturbations (GSUAP) to suppress inference accuracy when valid credentials are absent.

To rigorously benchmark the neutralizing potency of GSUAP, we conduct a comprehensive evaluation on three benchmark datasets: CIFAR-10~\cite{krizhevsky2009learning}, MNIST~\cite{lecun1998gradient}, and SVHN~\cite{netzer2011reading}. We construct two comparative baselines by adapting alternative neutralizing strategies: \textbf{Gaussian Noise} and \textbf{Sparse UAP (SUAP)}. SUAP is formulated by applying a sparsity mask to a standard UAP: $\Delta_{sparse} = \mathcal{M} \odot \Delta$. To ensure a fair comparison, all three strategies utilize the identical sparsity mask $\mathcal{M}$ generated by our \texttt{Weight-Gradient} selector, and their perturbation magnitudes are constrained to the same budget ($\|\epsilon\|_\infty=2$, applies to the normalized feature space after Batch Normalization). We integrate \textit{LymphNode} into a diverse suite of architectures to verify broad applicability, including standard residual networks (ResNet-18, ResNet-50~\cite{he2016deep}), vision transformers (ViT-Tiny, ViT-Small~\cite{dosovitskiy2021image}), and classic architectures (AlexNet~\cite{krizhevsky2012imagenet}, DenseNet~\cite{huang2017densely}). For each dataset, we randomly sample 2000 images, add watermark to 1000 of them as authorized input, while the rest as unauthorized input. We quantify effectiveness via the \textit{Unauthorized Accuracy} (lower is better) to measure the suppression effect, and the \textit{VIP Accuracy} (higher is better) to demonstrate the extent of performance preservation. The quantitative results are presented in Table~\ref{tab:full_comparison_gap_uniform}.

The quantitative results demonstrate that GSUAP universally outperforms baselines, evidenced by its ability to suppress ResNet-18 accuracy on CIFAR-10 to 13.6\% at a 60\% ratio, whereas Gaussian noise fails (85.4\%). This disparity confirms that structural corruption alone is insufficient, highlighting the necessity of gradient-guided semantic destruction for neutralizing robust models. We further observe that simpler datasets and densely connected architectures like DenseNet exhibit inherent fragility to perturbations due to error propagation, whereas the resilience of ResNet models validates the requirement for targeted adversarial interference. Throughout these configurations, authorized inputs maintain optimal accuracy, ensuring no performance penalty for legitimate users.

These observations validate the effectiveness of the \textit{LymphNode} framework. The consistent superiority of GSUAP establishes a reliable protection boundary that stochastic baselines cannot achieve, particularly on robust architectures. Simultaneously, the preservation of authorized fidelity confirms the precision of our feature-domain verification mechanism, which effectively decouples authorized flows from the noise injection path. In summary, \textit{LymphNode} provides a robust paradigm for active model protection, reconciling strict unauthorized lockout with seamless authorized access.

\subsection{Neutralization Efficiency Analysis}
\label{sec:efficiency}

\begin{figure*}[!t]
\centering
\includegraphics[width=0.8\textwidth]{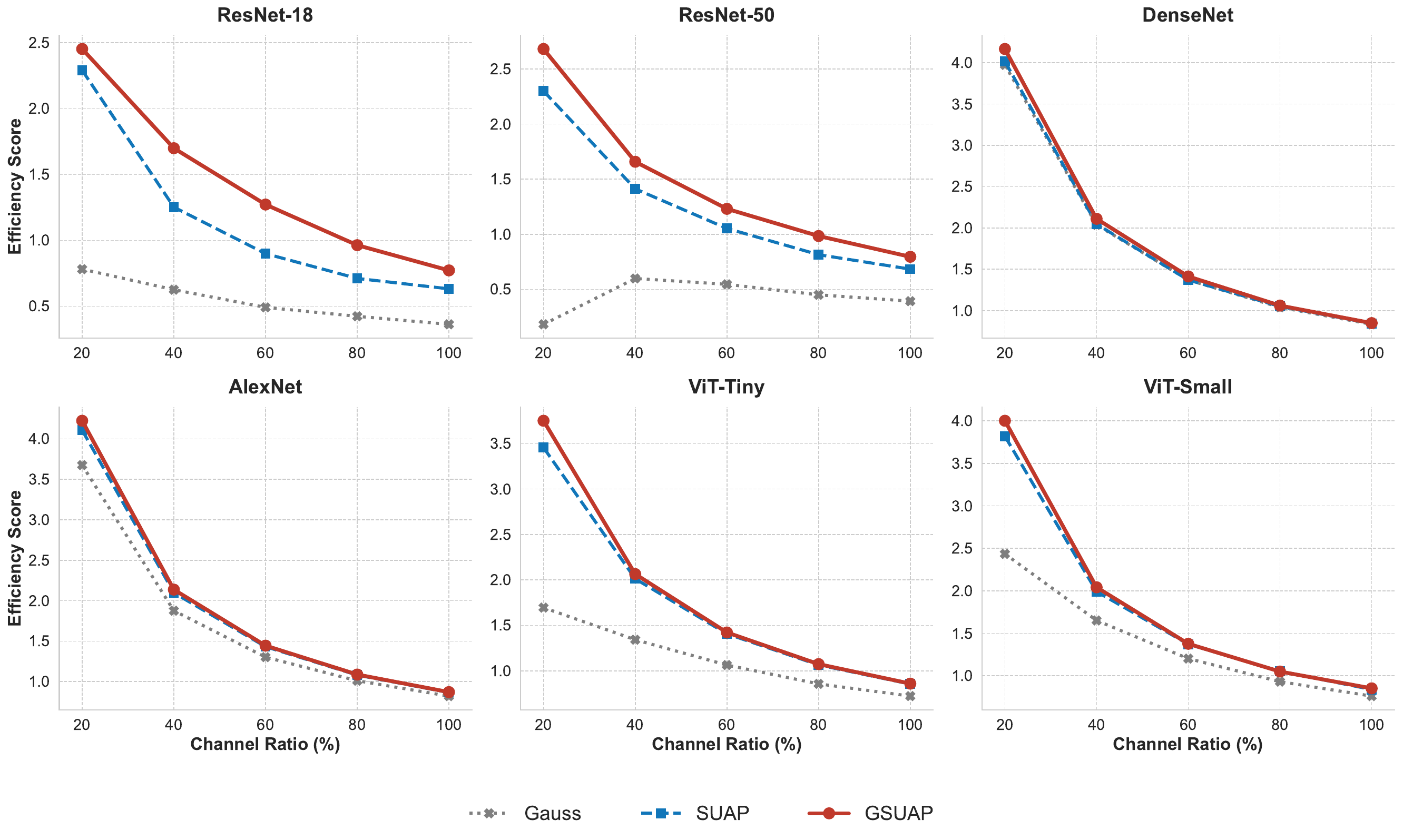}
\caption{Neutralization Efficiency.}
\label{fig:efficiency}
\end{figure*}

While the absolute performance drop is a critical measure of security, a practical protection mechanism must also be efficient, achieving maximum security with minimal structural modification. To rigorously quantify the return on investment for each injected noise channel, we introduce the Neutralization Efficiency metric $E$. Formally, defined by the performance gap between authorized and unauthorized users normalized by the sparsity ratio of modified channels $\rho$, the metric is calculated as:
\begin{equation}
    E = \frac{\mathcal{A}_{auth} - \mathcal{A}_{unauth}}{\rho}
\end{equation}
where $\mathcal{A}_{auth}$ and $\mathcal{A}_{unauth}$ represent the classification accuracy for authorized and unauthorized (Normal) inputs, respectively. Mathematically, this represents the marginal neutralization utility, quantifying how much protection gain is achieved per unit of channel resource consumed. To provide a holistic view of architectural behaviors that is not biased by specific data characteristics, we compute the efficiency scores for each model by averaging the results across all three datasets (CIFAR-10, MNIST, and SVHN).

As illustrated in Figure~\ref{fig:efficiency}, the efficiency trajectories across six architectures reveal distinct behavioral patterns governed by both perturbation strategy and model topology. Universally, the efficiency curves exhibit a monotonic decay as the channel ratio increases, a phenomenon consistent with the law of diminishing returns where the initial set of critical channels contributes most significantly to the decision boundary. Crucially, GSUAP consistently establishes a superior efficiency frontier compared to Gaussian noise and SUAP, particularly on robust architectures like ResNet and ViT. For instance, at a sparsity ratio of 20\% on ResNet-18, GSUAP achieves an efficiency score of nearly 2.5, significantly outperforming Gaussian noise ($<0.8$). This empirical evidence confirms that GSUAP does not merely introduce stochastic noise but precisely targets the model's semantic vulnerabilities. We further observe a divergence based on architectural resilience: for densely connected structures like DenseNet and legacy models like AlexNet, the efficiency curves of all three methods nearly collapse into a single trajectory, corroborating the ``error avalanche" effect where inherent architectural fragility renders the sophistication of the noise generation secondary to the perturbation itself. Conversely, for modern architectures possessing structural redundancy, such as ResNets and ViTs, the performance gap between GSUAP and the baselines widens significantly.

Synthesizing these observations, we conclude that the efficiency of gradient-guided adversarial perturbations is fundamentally superior to stochastic interference, particularly when protecting robust models. The pronounced ``efficiency gap" observed on resilient architectures (e.g., ResNet) highlights that while random noise may suffice for fragile models, the precise, adversarial nature of GSUAP is indispensable for neutralizing robust networks under strict resource constraints. Consequently, these results validate the core design premise of \textit{LymphNode}: by leveraging GSUAP, our framework achieves substantial model degradation with minimal feature modification. This high operational efficiency ensures that the protection mechanism imposes the lowest possible structural cost while maintaining a robust lockout against unauthorized access.

\subsection{System Performance Analysis}
\label{sec:overhead}

For any active protection mechanism to be viable in real-world deployments—particularly on resource-constrained edge devices—it must impose minimal computational burden. In this section, we rigorously quantify the system overhead introduced by the \textit{LymphNode} plugin.

In detail, we benchmark the inference performance on an NVIDIA GeForce RTX 4060 Laptop GPU. We compare the original clean models against their protected counterparts across five key metrics: parameter count (Storage), theoretical FLOPs (Computational Complexity), inference latency (Batch Size=1), throughput (Batch Size=128), and peak memory usage. We select ResNet-18 and ViT-Tiny as representative architectures for CNNs and Transformers, respectively. Test data is sampled from watermarked CIFAR-10 dataset.

\begin{table}[ht]
    \centering
    \caption{System overhead analysis.}
    \label{tab:overhead}
    \small
    \setlength{\tabcolsep}{2.5pt} 
    \renewcommand{\arraystretch}{1.15}
    
    \begin{tabular}{l | ccc | ccc}
        \toprule
        \multirow{2}{*}{\textbf{Metric}} & \multicolumn{3}{c|}{\textbf{ResNet-18}} & \multicolumn{3}{c}{\textbf{ViT-Tiny}} \\
        \cmidrule(lr){2-4} \cmidrule(lr){5-7}
         & Clean & Prot. & \textbf{Cost} & Clean & Prot. & \textbf{Cost} \\
        \midrule
        Params (M)      & 11.17 & 11.17 & +0.0\%   & 1.80  & 1.80  & +0.0\% \\
        FLOPs (G)       & 0.558 & 0.558 & $\approx 0$ & 0.118 & 0.118 & $\approx 0$ \\
        Latency (ms)    & 1.70  & 2.74  & \textbf{+1.0} & 1.54  & 2.60  & \textbf{+1.1} \\
        Throughput (/s) & 6893  & 5875  & -14.8\%  & 17175 & 16023 & -6.7\% \\
        Memory (MB)     & 93.2  & 109.2 & +17.2\%  & 33.7  & 35.2  & +4.5\% \\
        \bottomrule
    \end{tabular}
\end{table}
\textbf{Results Analysis.}
Table~\ref{tab:overhead} summarizes the results. 
First, regarding theoretical complexity, the plugin introduces negligible overhead in terms of parameters and FLOPs ($<0.01\%$). This confirms that the element-wise operations (LSB extraction and noise injection) are computationally lightweight compared to the backbone's convolutional layers.
Second, regarding runtime performance, the throughput reduction is modest ($-6.7\%$ to $-14.8\%$), maintaining high efficiency for batch processing.
Finally, for real-time latency, although the relative increase appears significant due to the extremely low baseline of CIFAR models, the absolute latency penalty is consistently around 1.0 ms. In practical scenarios, such as video processing at 30 FPS (requiring $<33$ ms per frame), a total inference time of $\approx 2.7$ ms is well within the operational envelope. This confirms that \textit{LymphNode} achieves strong active protection without compromising system responsiveness.

\subsection{Channel Selection Evaluation}
\label{sec:exp_ablation_selection}

\begin{figure*}[!t]
\centering
\includegraphics[width=0.99\textwidth]{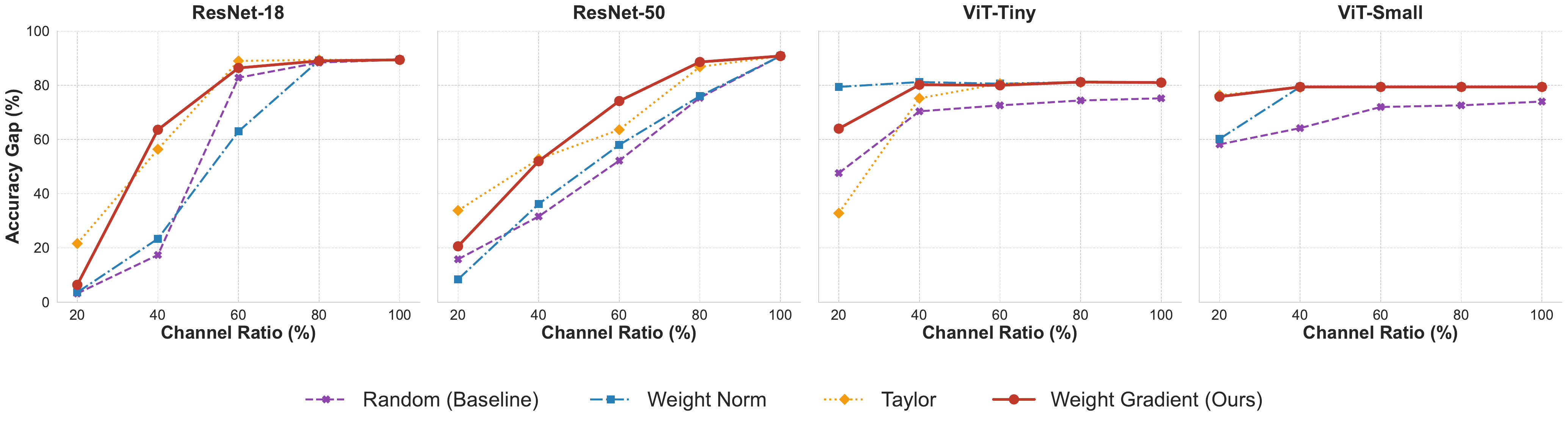}
\caption{Ablation study for selector.}
\label{fig:selection_ablation}
\end{figure*}

To validate the channel selection strategy within \textit{LymphNode}, we conduct an ablation study on the CIFAR-10 dataset to determine the most effective criterion for identifying critical features. We benchmark our \textbf{Weight Gradient} approach against three established saliency metrics derived from model pruning literature: Random selection (baseline), \textbf{Weight Norm}~\cite{li2016pruning}, and \textbf{Taylor Expansion}~\cite{molchanov2019importance}. To ensure a strictly fair comparison, we apply these criteria to generate the sparsity mask while keeping the noise injection method (GSUAP) constant. The evaluation focuses on four representative architectures: ResNet-18, ResNet-50, ViT-Tiny, and ViT-Small. These models are selected because their inherent structural redundancy offers high resilience to interference, providing a rigorous testing ground to distinguish the efficacy of different selection strategies. All necessary selection metrics, such as gradients and weight magnitudes, are computed directly from the corresponding clean models pre-trained on CIFAR-10. We quantify the performance using the Accuracy Gap (defined as $\mathcal{A}_{auth} - \mathcal{A}_{unauth}$), where a wider gap indicates a more efficient selection method that achieves greater performance degradation for unauthorized users while maintaining the same sparse footprint. The results are presented in Figure~\ref{fig:selection_ablation}.

The results provide a clear justification for our design choice. First, we observe that perturbing a specific proportion of critical channels (e.g., 60\% on ResNet and 40\% on ViT) yields neutralization effects comparable to full-channel perturbation, verifying that targeted noise injection significantly enhances efficiency without compromising potency. Universally, the selection strategy proves decisive; both gradient-based methods significantly outperform the magnitude-based and random baselines, confirming that sensitivity to the loss function is a superior proxy for channel importance compared to static magnitude heuristics. Crucially, between the top performers, our \textbf{Weight Gradient} method demonstrates superior stability and overall efficiency. While \textbf{Taylor Expansion} is competitive at extreme sparsity levels for smaller models, it lacks consistency in the practical 40\%--80\% range. For instance, on ResNet-50 at a 60\% ratio, our method achieves an Accuracy Gap of approximately 75\%, substantially surpassing the 52\% achieved by Taylor Expansion. Furthermore, on ViT-Tiny, our approach maintains a distinct performance lead across all ratios. Consequently, we adopt the weight gradient criterion as the default selector for our framework due to its robust trade-off between sparsity and neutralization capability.
\subsection{Impact of Noise Scale}
\label{sec:ablation_noise}

A distinct advantage of our plugin-based architecture is the ability to dynamically regulate the severity of performance degradation. Unlike weight-embedded backdoors that typically offer a binary ``all-or-nothing" outcome, \textit{LymphNode} allows the model provider to tune the intensity of the injected Generalized Sparse Universal Adversarial Perturbation (GSUAP) to achieve granular access control. To demonstrate this capability, we conduct a representative case study using ResNet-18 on the CIFAR-10 dataset. We fix the sparse channel selection ratio at $40\%$ and systematically vary the noise scale factor $\lambda$ from $0.0$ to $2.0$ with a step size of $0.1$. Here, $\lambda=0$ represents an unprotected model, while higher values indicate stronger interference. We measure the inference accuracy for both authorized users and unauthorized users.

\begin{figure}[!t]
    \centering
    \includegraphics[width=0.8\linewidth]{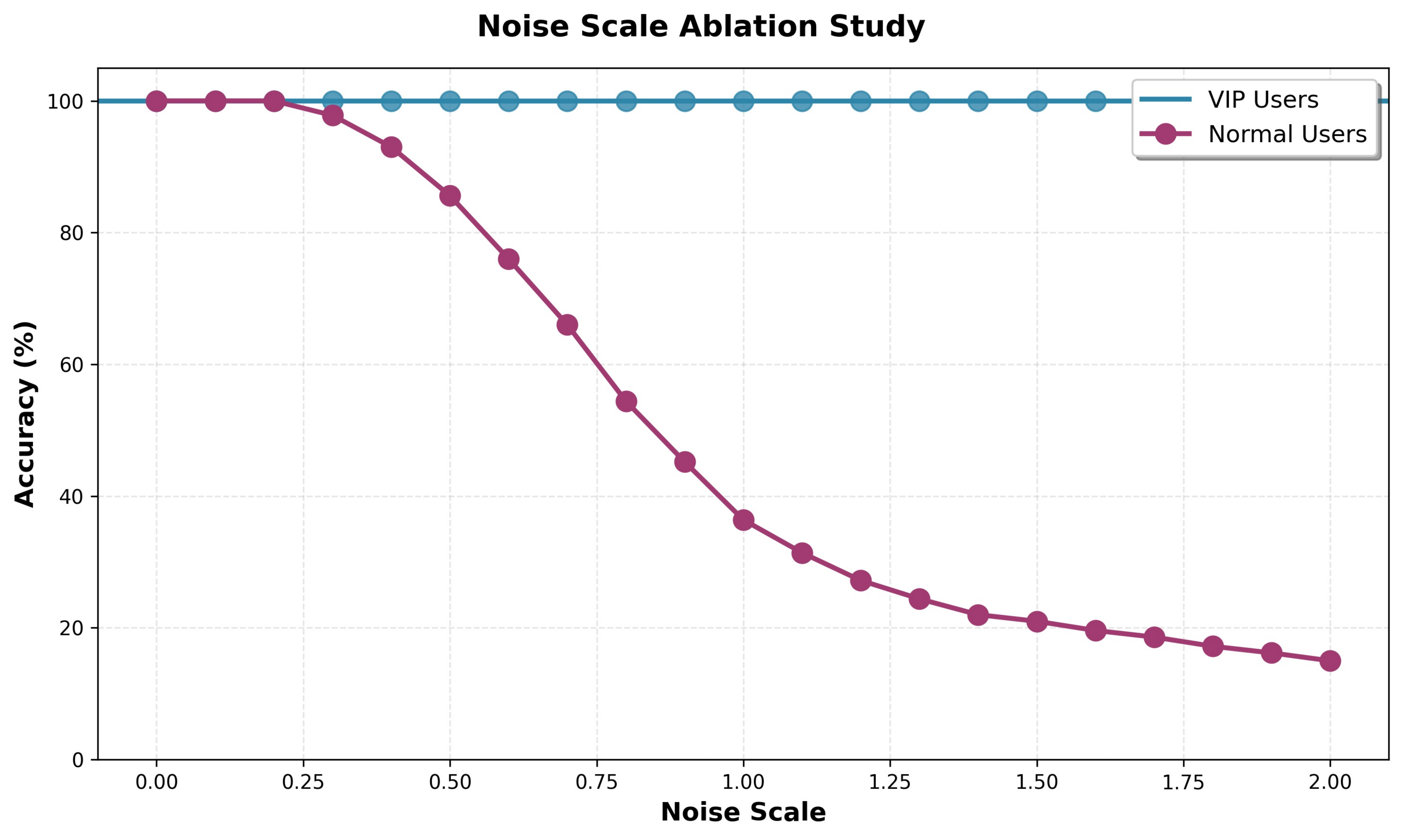} 
    \caption{Impact of Noise Scale ($\lambda$) on model performance (ResNet-18 on CIFAR-10).}
    \label{fig:noise_ablation}
\end{figure}

As illustrated in Figure~\ref{fig:noise_ablation}, the performance trajectories reveal a contrast between user groups. The authorized User accuracy (blue line) remains robustly stable at the optimal level ($\approx 99.8\%$) across the entire range of $\lambda$, confirming that our identity verification module reliably triggers the antidote mechanism. Conversely, the performance for unauthorized users (red line) exhibits a smooth, monotonic decline as $\lambda$ increases. This experiment validates the controllability of our framework, demonstrating that \textit{LymphNode} enables flexible deployment strategies ranging from subtle quality degradation to complete functionality lockout without the need to retrain the model.

\subsection{Real Attack Scenarios}
\label{Sec:real attack exp}

To directly validate the defense against practical IP threats, we evaluate \textit{LymphNode} under two representative attack pipelines with oracle-access conditions consistent with Sec.~\ref{sec:threat_model}. We include PP~\cite{orekondy2020prediction} and CIP~\cite{zhang2023cip} as baselines. Since both methods are originally designed for cloud-API deployment with runtime query interposition, we preserve their core defense logic (per-query gradient perturbation for PP, reliability-based poisoning for CIP), remove the query-detection components that require server-side control, and activate their defense on all queries unconditionally. This adaptation is strictly favorable to the baselines, as it eliminates any detection failure and ensures 100\% defense activation.

\noindent\textit{Model Extraction.} We adopt the KnockoffNets~\cite{orekondy19knockoff} protocol, where an adversary trains a substitute model from input--output pairs collected from the victim. The target is ResNet-18 on CIFAR-10 with channel ratio 60\% and noise scale 2.0. The query budget varies from 1{,}000 to 50{,}000; results are reported in Table~\ref{tab:extraction}.

\begin{table}[ht]

    \centering
    \caption{KnockoffNet extraction attack on CIFAR-10.}
    \label{tab:extraction}
    \small
    \renewcommand{\arraystretch}{1.25}
    \setlength{\tabcolsep}{6pt}
    
    \begin{tabular}{lcccc}
        \toprule
        \textbf{Budget} & \textbf{No Def.} & \textbf{PP~\cite{orekondy2020prediction}} & \textbf{CIP~\cite{zhang2023cip}} & \textbf{Ours} \\
        \midrule
        Victim Acc. & 92.54 & 92.54 & 92.54 & 92.54 \\
        \midrule
        1,000   & 42.88 & 25.26 & 28.44 & \textbf{13.71} \\
        5,000   & 56.25 & 32.68 & 35.17 & \textbf{14.76} \\
        10,000  & 66.16 & 38.24 & 42.81 & \textbf{14.68} \\
        50,000  & 85.24 & 45.66 & 48.25 & \textbf{15.28} \\
        \bottomrule
    \end{tabular}
    
    \vspace{4pt}
    \parbox{\columnwidth}{\footnotesize Ours: non-VIP model accuracy = 16.47\%, output entropy = 1.35 (uniform = 2.30).}
\end{table}

Without defense, surrogate accuracy reaches 85.24\% at 50K queries. PP and CIP reduce this to 45.66\% and 48.25\%, yet surrogate accuracy continues climbing with budget, indicating persistent information leakage through perturbed outputs. \textit{LymphNode} suppresses surrogate accuracy to 13.71--15.28\% across all budgets---near random guessing (10\%)---with no improvement as budget scales by 50$\times$, confirming that feature-space corruption renders collected soft labels uninformative for distillation.

\noindent\textit{Model Inversion.} We adopt the GMI framework~\cite{zhang2020secret}, where a DCGAN trained on public face images is optimized in latent space to reconstruct private training samples of a CelebA classifier. To comply with the oracle-access constraint, we replace white-box gradients with Natural Evolution Strategies (NES), yielding a fully black-box variant. The attack targets 50 randomly selected identities. Attack success is measured by an independent evaluator: Acc-1/5 (lower is better) indicates how often reconstructions match the target identity; KNN Distance (higher is better) quantifies feature-space divergence from real samples. Results appear in Table~\ref{tab:inversion}.

\begin{table}[ht]
    \centering
    \caption{GMI model inversion attack results on CelebA (50 identities).}
    \label{tab:inversion}
    \small
    \renewcommand{\arraystretch}{1.2}
    \setlength{\tabcolsep}{6pt}

    \begin{tabular}{lcccc}
        \toprule
        Defense & Acc-1$\downarrow$ & Acc-5$\downarrow$ & KNN Dist$\uparrow$ & Latency \\
        \midrule
        None (Clean)  & 82.71\% & 91.32\% & 5.57 & 0.082ms \\
        CIP           & 44.30\% & 48.23\% & 6.45 & \underline{0.110ms} \\
        PP            &  \textbf{3.02\%} &  \textbf{8.77\%} & \textbf{7.71} & 0.308ms \\
        Ours          &  \underline{4.17\%} &  \underline{11.22\%} & \underline{7.51} & \textbf{0.088ms} \\
        \bottomrule
    \end{tabular}
\end{table}

CIP provides limited defense (Acc-1 = 44.30\%). PP achieves the strongest inversion protection (Acc-1 = 3.02\%) but performs per-query PGD optimization, increasing latency from 0.082\,ms to 0.308\,ms---a 275\% overhead. CIP incurs a moderate 34\% increase (0.110\,ms) via per-query reliability scoring. \textit{LymphNode} achieves comparable protection (Acc-1 = 4.17\%, Acc-5 = 11.22\%, both near random guessing) with only 7.3\% latency increase (0.088\,ms), as GSUAP is a pre-computed static tensor requiring only a constant-cost element-wise addition. For throughput-sensitive edge applications such as real-time video analytics, this constant-overhead property is essential---any per-query computation directly erodes the available compute budget on resource-constrained devices.

Across both attacks, \textit{LymphNode} matches or exceeds the strongest baseline in protection while adding negligible inference cost, making it uniquely suited for latency-constrained edge deployment.

\section{Data Adaptivity}
\label{sec:data adaptivity}

In this section, we use two experiments to highlight that \texttt{LymphNode} has strong dataset adaptivity. We validate that our plugin framework can achieve performance control with limited data sampling from a similar domain.

\subsection{Dataset Size}
\label{sec:dataset size exp}

\begin{figure*}[ht]
\centering
\includegraphics[width=0.99\textwidth]{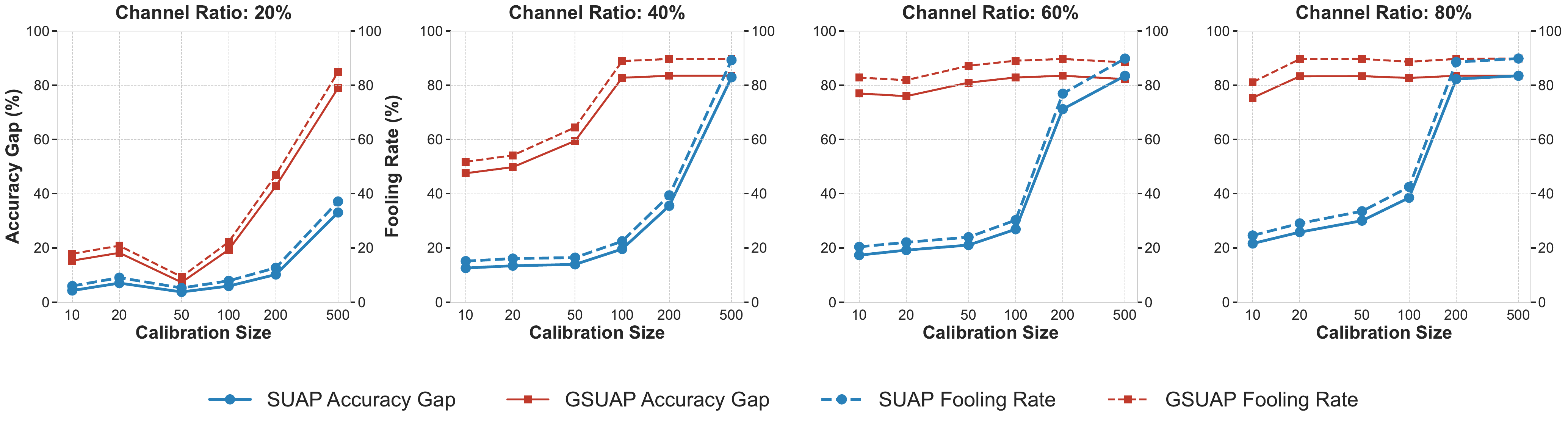}
\caption{effectiveness with different dataset size.}
\label{fig:training_data_size}
\end{figure*}

Here, we evaluate the influence of calibration data size on the effectiveness of \textit{LymphNode}. Using the ResNet-18 architecture on CIFAR-10 as a testbed, we optimize the internal GSUAP module using varying subsets of the training data, ranging from 10 to 500 samples, across channel ratios from $20\%$ to $100\%$. It is crucial to note that the maximum budget of 500 samples constitutes merely \textbf{1\%} of the original dataset, simulating an extremely low-resource setup. We evaluate performance on the test set using the \textbf{Accuracy Gap} and the \textbf{Fooling Rate}. The Fooling Rate measures the proportion of clean images that are misclassified by the target network solely due to the injected perturbation, defined as:
\begin{equation}
\text{Fooling Rate} = \frac{N_{\text{misclassified\_with\_GSUAP}}}{N_{\text{correctly\_classified\_clean}}}
\end{equation}
In our visualization, solid and dashed lines distinguish these two metrics. We benchmark against the SUAP baseline for comparison, while Gaussian noise is excluded as prior experiments (Sec.~\ref{sec:exp_effectiveness}) have already demonstrated its insufficient neutralizing capability. The results are presented in Figure~\ref{fig:training_data_size}.

The results in Figure~\ref{fig:training_data_size} demonstrate the superior data efficiency of \textit{LymphNode}. We observe a significant performance gap between our method and the SUAP baseline. GSUAP exhibits a rapid convergence trajectory, establishing a robust lockout with minimal data. Specifically, at channel ratios of $40\%$ and above, GSUAP achieves a high Accuracy Gap and Fooling Rate with as few as 50 to 100 samples. Furthermore, at higher channel ratios ($60\%$ and $80\%$), the protection capability saturates quickly, reaching peak performance with a negligible data budget (approx. 50 samples). These observations confirm that \textit{LymphNode} entails low setup costs and is highly practical for real-world deployment.

\subsection{Domain-Adaptivity Analysis}

\begin{table}[ht]
\centering
\caption{Cross-Dataset Effectiveness: GSUAP generated from CIFAR-10 (Source) applied to models trained on distinct Target Domains.}
\label{tab:transfer}
\renewcommand{\arraystretch}{1.2}
\resizebox{\columnwidth}{!}{
    \setlength{\tabcolsep}{4pt}
    \begin{tabular}{c@{\hspace{30pt}}cc@{\hspace{20pt}}cc@{\hspace{20pt}}cc}
    \toprule
    \multirow{2}{*}{\textbf{Ratio}} & 
    \multicolumn{2}{c}{\textbf{CIFAR-10 (Source)}} & 
    \multicolumn{2}{c}{\textbf{CIFAR-100 (Target)}} & 
    \multicolumn{2}{c}{\textbf{STL-10 (Target)}} \\
    \cmidrule(lr){2-3} \cmidrule(lr){4-5} \cmidrule(lr){6-7}
     & \textbf{Gap} & \textbf{FR} & \textbf{Gap} & \textbf{FR} & \textbf{Gap} & \textbf{FR} \\
    \midrule
    0.2 & 19.34 & 22.22 & 40.82 & 43.77 & 14.70 & 20.17 \\
    0.4 & 82.79 & 88.94 & 69.98 & 64.13 & 38.35 & 42.58 \\
    0.6 & 82.89 & 89.07 & 70.66 & 75.02 & 46.06 & 50.20 \\
    0.8 & 82.71 & 88.68 & 70.65 & 75.14 & 51.31 & 55.11 \\
    \bottomrule
    \multicolumn{7}{l}{\footnotesize All metrics reported in percentage (\%). \textbf{Gap}: Accuracy Gap; \textbf{FR}: Fooling Rate.}
    \end{tabular}
}
\end{table}

In practical scenarios, the model provider might need to protect a model without accessing its original training set. To evaluate whether \textit{LymphNode} can be effectively deployed using accessible proxy datasets, we conducted transferability experiments by optimizing GSUAP on the CIFAR-10 dataset (Source) and deploying the generated perturbation to neutralize ResNet-18 models trained on different target domains: CIFAR-100 and STL-10. We selected these targets to isolate domain characteristics; CIFAR-100 shares the same resolution ($32\times32$) as the source but differs in semantic granularity, while STL-10 introduces a larger domain shift with disjoint class definitions and higher resolution ($96\times96$). For evaluation, we randomly sampled 500 clean images from each target dataset and applied the source-derived perturbation, quantifying efficacy via the Accuracy Gap. The results in Table~\ref{tab:transfer} reveal remarkable adaptability, particularly when the domains share statistical similarities. Specifically, the CIFAR-10 derived perturbation achieves a control efficacy on the CIFAR-100 target that closely trails the oracle setting at higher channel ratios. Notably, at the lowest sparsity (0.2), the cross-domain attack on CIFAR-100 ($40.82\%$ Gap) significantly outperforms the source attack on CIFAR-10 ($19.34\%$ Gap). We attribute this to the inherent fragility of the fine-grained CIFAR-100 model, where the coarse-grained adversarial patterns learned from CIFAR-10 prove disruptively potent against the more complex decision boundaries. While efficacy naturally degrades on the more distant STL-10 domain due to resolution mismatches, functional control remains intact.

Synthesizing these observations, we conclude that access to the original training dataset is not a prerequisite for \textit{LymphNode}. The strong transferability demonstrates that as long as the surrogate dataset aligns with the target in basic visual statistics, the \textit{LymphNode} framework can establish a robust lockout mechanism. This finding is pivotal for practical application, as it significantly lowers the barrier for adoption. It validates that model owners can reliably initialize the protection mechanism using publicly accessible surrogate datasets, thereby securing their intellectual property even in scenarios where the original training data is private, lost, or computationally too expensive to process.

\section{Robustness Evaluation}
\label{sec:robustness}

We strictly evaluate the dependability of \textit{LymphNode} against a knowledgeable adversary whose objective is either to forge valid credentials or to physically remove the protection mechanism. In this evaluation, we assume a ``deployment-ready" threat model where the underlying system integrity is secured, limiting the adversary to adaptive attacks involving traffic interception, credential fabrication, and model modification. To demonstrate the superiority of our plugin-based architecture over traditional parameter-based protections, we implement two representative baselines: BadNets~\cite{gu2017badnets} and Blended Attack~\cite{chen2017targeted}. We re-purpose these backdoor attacks as baseline access control mechanisms by training models to yield high-fidelity inference only upon detecting specific triggers—a visible $3\times3$ patch for BadNets and an invisible static noise pattern ($\alpha=0.1$) for Blended Attack. This setup allows for a rigorous comparative analysis. Our evaluation comprehensively covers four critical dimensions: watermark imperceptibility, robustness against credential forgery, resilience to fine-tuning attacks, and stability under lossy compression and pruning.

\subsection{Watermark Imperceptibility Analysis}

\begin{table}[ht]
    \centering
    \caption{Quantitative comparison of watermark imperceptibility on CIFAR-10. $\uparrow$ indicates higher is better; $\downarrow$ indicates lower is better.}
    \label{tab:imperceptibility_stats}
    \small 
    \renewcommand{\arraystretch}{1.25} 
    \setlength{\tabcolsep}{10pt} 
    
    \begin{tabular}{lccc}
        \toprule
        \textbf{Method} & \textbf{PSNR (dB)} $\uparrow$ & \textbf{SSIM} $\uparrow$ & \textbf{LPIPS} $\downarrow$ \\
        \midrule
        BadNets  & $28.3502 $ & $0.9370 $ & $0.0472$ \\
        Blended  & $33.9331 $ & $0.8245 $ & $0.0218 $ \\
        \midrule
        \textbf{LymphNode} & $\mathbf{56.6411 }$ & $\mathbf{0.9990 }$ & $\mathbf{0.0011 }$ \\
        \bottomrule
    \end{tabular}
\end{table}
To ensure authorization credentials remain undetectable to human or automated inspection, we evaluate visual stealthiness on 1,000 randomly sampled CIFAR-10 image pairs using PSNR, SSIM, and Learned Perceptual Image Patch Similarity (LPIPS). While standard metrics focus on pixel-level differences, LPIPS measures perceptual distance in deep feature space, serving as a critical indicator for high-frequency artifacts that might escape traditional metrics.

The quantitative results in Table~\ref{tab:imperceptibility_stats} highlight the superior fidelity of our approach. The baselines exhibit clear limitations: BadNets suffers from low PSNR (28.35 dB) due to concentrated corruption, while the Blended Attack degrades structural quality (SSIM 0.8245) through global noise injection. In stark contrast, LymphNode achieves near-perfect imperceptibility across all metrics, attaining a PSNR of 56.64 dB and an LPIPS score of 0.0011—an order of magnitude superior to the baselines. This stealthiness is fundamentally attributed to our precision-targeted embedding strategy, which targets deep bit-depths ($s \ge 5$) to restrict feature modifications to the level of numerical quantization noise, thereby preserving both pixel statistics and deep perceptual integrity.

\subsection{Robustness against Credential Forgery}

\begin{table}[ht]
    \centering
    \caption{Robustness against Credential Forgery attacks. We report the \textit{Forgery Success Rate} (success rate of forged credentials) on the protected models. A lower accuracy indicates higher robustness against forgery.}
    \label{tab:forgery_results}
    \renewcommand{\arraystretch}{1.2} 
    \begin{tabular}{llc}
        \toprule
        \textbf{Target Model} & \textbf{Attack Method} & \textbf{Forgery Success Rate. (\%)} \\
        \midrule
        \multirow{2}{*}{BadNets (Baseline)} & Linear Est. & 100.0 \\
                                            & U-Net       & 100.0 \\
        \midrule
        \multirow{2}{*}{Blended (Baseline)} & Linear Est. & 0.0 \\
                                            & U-Net       & 23.8 \\
        \midrule
        \multirow{2}{*}{\textbf{Ours (LymphNode)}} & Linear Est. & \textbf{0.0} \\
                                                   & U-Net       & \textbf{0.0} \\
        \bottomrule
    \end{tabular}
\end{table}

To evaluate robustness against credential forgery, we simulate an adversary who attempts to reverse-engineer the authorization watermark using 500 intercepted pairs of unauthorized and authorized images. We benchmark our method against the Blended-based model as a primary competitor due to its use of invisible static triggers. The evaluation employs two approximation attacks targeting the mathematical properties of the watermark: Linear Residual Estimation, which assumes a static additive signal derived by averaging residuals to suppress image-specific variations, and a Deep Mapping Attack via U-Net, which treats forgery as a supervised image-to-image translation task minimizing pixel-wise reconstruction loss. We quantify robustness using the Forgery Success Rate on forged images, hypothesizing that while baselines relying on continuous pixel-domain modifications are susceptible to approximation, our discrete feature-bit alignments will resist such replication.

The quantitative results in Table~\ref{tab:forgery_results} validate this hypothesis, revealing distinct vulnerability patterns. The BadNets baseline yields a 100\% forgery success rate due to its simple trigger, while the Blended Attack remains vulnerable to U-Net (23.8\% success rate), indicating that its reliance on continuous feature activations allows for approximation by generative models given sufficient data. In stark contrast, \textit{LymphNode} achieves absolute robustness with a 0.0\% success rate against all attempts. This superior security is attributed to the fundamental verification discontinuity inherent in our design; while generative models like U-Net inherently produce smooth outputs with microscopic floating-point residuals, these deviations disrupt the precise quantization logic required for our LSB-based verification. Consequently, our discrete feature-domain mechanism establishes a mathematical barrier that renders forged credentials invalid, proving intractable for standard continuous approximation algorithms.

\subsection{Robustness against Fine-tuning Attacks}

\begin{figure}[t] 
    \centering
    \includegraphics[width=0.95\linewidth]{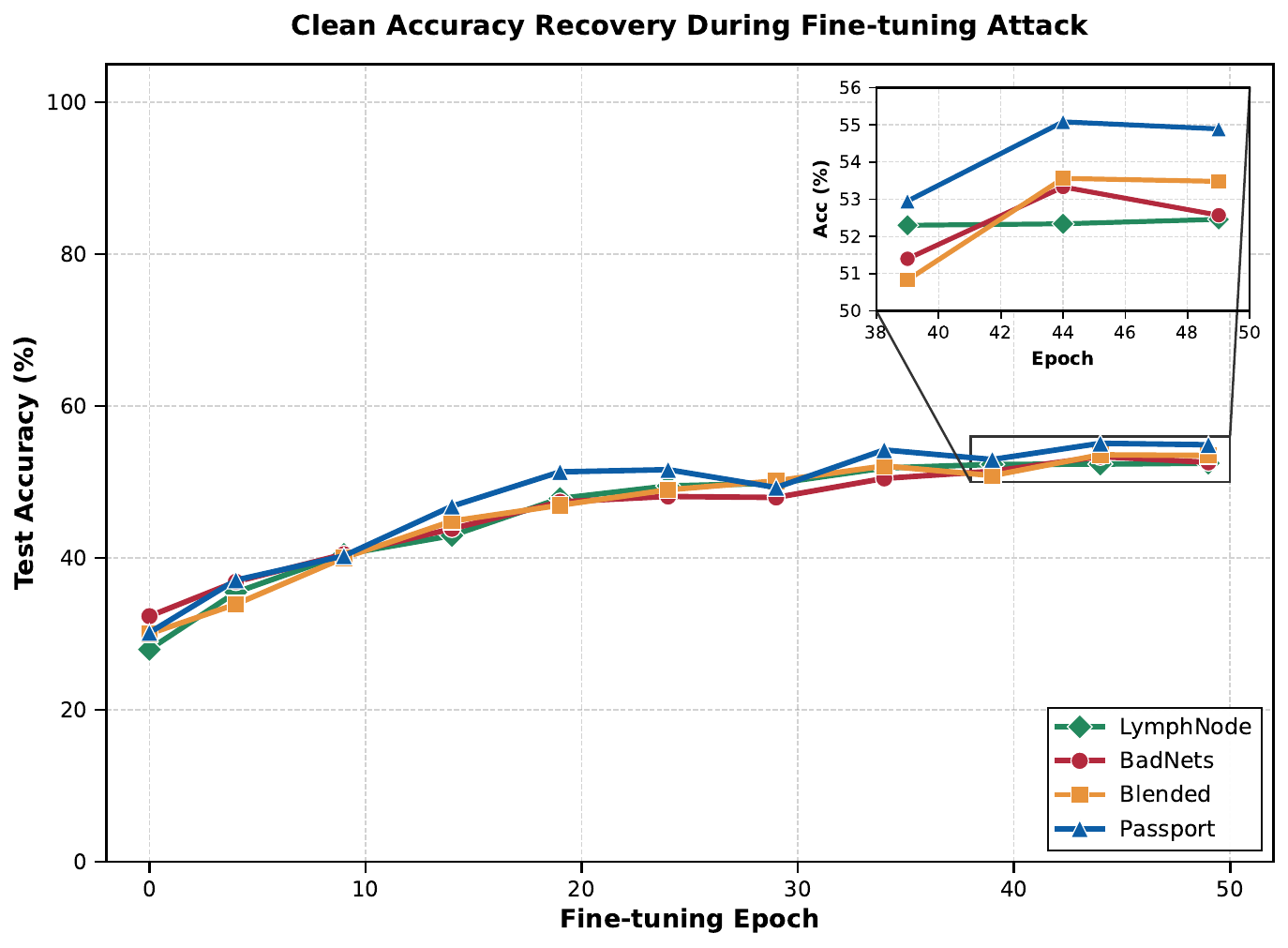}
    \caption{Robustness against fine-tuning attacks. The trajectories illustrate the Clean Accuracy recovery over 50 epochs. }
    \label{fig:hijacking_comparison}
\end{figure}

Beyond credential forgery, an adversary possessing the model artifacts might attempt to attack the protection mechanism through fine-tuning. We simulate a gray-box scenario where the adversary has access to the model parameters and clean training data ($10\%$), but lacks knowledge of the specific plugin logic. To evaluate resilience, we benchmark against BadNets, Blended Attack, and Passport~\cite{fan2022deepipr} under identical conditions. The adversary employs standard Stochastic Gradient Descent (SGD) with a learning rate of $0.01$ to fine-tune the protected models. The fine-tuning process is conducted for 50 epochs using the clean CIFAR-10 dataset.

Figure~\ref{fig:hijacking_comparison} illustrates the accuracy recovery trajectories. We observe that after 50 epochs, the clean accuracy of the \textit{LymphNode}-protected model recovers to $52.46\%$, which remains significantly below the original fidelity ($\approx 95\%$). This final accuracy is comparable to that of the structural baseline (Passport, $54.89\%$) and the data-poisoning baselines ($\approx 53\%$). These results indicate that \textit{LymphNode}, despite being a post-hoc plugin, achieves a level of resistance against fine-tuning that is on par with heavyweight structural defenses like Passport, effectively preventing the restoration of model utility under standard fine-tuning protocols.

\subsection{Resilience to Lossy Compression}
In practical deployment scenarios, input images frequently undergo lossy compression during transmission, such as JPEG encoding, which quantizes high-frequency DCT coefficients and inherently challenges the precision-based LSB verification mechanism. To evaluate the robustness of our system against such channel distortions, we measured the Authorization Success Rate (ASR) of credentials subjected to JPEG compression across varying quality factors. Recognizing the sensitivity of standard single-step embedding to aggressive quantization, we adopted an iterative embedding strategy that progressively refines pixel perturbations to survive the specific quantization matrix of the target quality level. 
\begin{table}[ht]
    \centering
    \caption{Authorization Success Rate (ASR) evolution using the Iterative Embedding Strategy~\cite{liu2022loneneuron}.}
    \label{tab:jpeg_iterations}
    \small
    \setlength{\tabcolsep}{6pt}
    \renewcommand{\arraystretch}{1.2}
    
    \begin{tabular}{c|ccccc}
        \toprule
        \multirow{2}{*}{\textbf{Quality ($Q$)}} & \multicolumn{5}{c}{\textbf{Iterations ($T$)}} \\
        \cline{2-6}
         & 10 & 20 & 30 & 40 & 50 \\
        \midrule
        80 & 15.6\% & 82.4\% & 96.5\% & 99.1\% & 99.8\% \\
        70 & 2.3\%  & 45.8\% & 81.2\% & 93.7\% & 98.4\% \\
        60 & 0.0\%  & 8.5\%  & 35.2\% & 62.8\% & 85.6\% \\
        \bottomrule
    \end{tabular}
\end{table}

The performance evolution, detailed in Table~\ref{tab:jpeg_iterations}, demonstrates the efficacy of this adaptive approach. While the survival rate is initially negligible under severe compression (e.g., quality factor 60), it improves dramatically to over 85\% after 50 iterations. These results confirm that \textit{LymphNode} can be effectively hardened to guarantee reliable access for authorized users even in bandwidth-constrained environments, successfully balancing strict security requirements with practical usability.

\subsection{Resistance to Model Pruning.} 
While we focus on fine-tuning, model pruning represents another potential removal vector. However, our framework inherently mitigates this threat through the design of the \textbf{Weight-Gradient} selector (evaluated in Sec.~\ref{sec:exp_ablation_selection}). As demonstrated in Fig.~\ref{fig:selection_ablation}, the protection mechanism is anchored to the channels with the highest gradient sensitivity—i.e., the most ``decision-critical" features of the network. Standard pruning algorithms, which eliminate redundant neurons~\cite{han2015learning}, would naturally bypass our plugin-associated channels. Conversely, an aggressive adversary attempting to prune these specific high-saliency channels would inadvertently destroy the model's primary classification capability before successfully removing the protection. This establishes a \textbf{structural coupling} where the survival of the security mechanism is tied to the utility of the model itself, rendering pruning attacks ineffective.

\section{Discussion and Future Work}
\label{sec:discussion}
While \textit{LymphNode} demonstrates efficient and robust performance, we acknowledge two specific limitations in the current implementation. First, as a software-level plugin, the defense relies on the integrity of the hosting runtime environment; an adversary with unrestricted write access to the model parameters could theoretically identify and bypass the protection logic. Future work can address this by coupling the plugin with lightweight code obfuscation or hardware-backed integrity checks to withstand physical tampering. Second, the current verification mechanism relies on fine-grained feature perturbations optimized for floating-point models. In scenarios employing aggressive quantization (e.g., INT8) for extreme compression, these subtle credentials may be distorted or lost. Future works can extend the framework to support quantization-aware embedding strategies compatible with low-precision arithmetic.

\section{Conclusion}
\label{sec:conclusion}
In this paper, we introduced \textit{LymphNode}, a novel post-hoc plugin framework designed to secure Deep Neural Networks against model extraction and inversion attacks. Addressing the critical vulnerability of unrestricted oracle access, the framework establishes an immunological ``default-deny'' checkpoint. It actively neutralizes model utility via Generalized Sparse Universal Adversarial Perturbations (GSUAP), effectively blocking the gradient estimation required for extraction, while transparently restoring fidelity for authorized users through a discrete feature-domain verification mechanism. Specifically, its ability to initialize robust protection using only public surrogate datasets—combined with a strictly constant $O(1)$ inference overhead—ensures that the framework can be seamlessly scaled and instantly deployed across heterogeneous, resource-constrained edge nodes without hardware-specific tuning.

Our comprehensive evaluation confirms that \textit{LymphNode} successfully reconciles rigorous security with operational practicality. We demonstrated that the protection is exceptionally data-efficient, establishing robust neutralization with as few as 50--100 calibration samples ($<1\%$ of training data) that need not originate from the original dataset, thus eliminating the dependency on sensitive private data required by prior art. From a system perspective, the plugin introduces negligible overhead ($\approx 1$ ms latency), ensuring viability for real-time applications. Furthermore, robustness analysis verifies that our credentials achieve near-perfect imperceptibility and exhibit strong resilience against adaptive threats, including generative forgery and fine-tuning. By decoupling IP protection from the constraints of model retraining, \textit{LymphNode} offers a scalable, deployment-ready solution for safeguarding high-value AI assets.


\bibliographystyle{IEEEtran}
\bibliography{refer}

@inproceedings{liu2022loneneuron,
  author    = {Liu, Zeyan and Li, Fengjun and Li, Zhu and Luo, Bo},
  title     = {LoneNeuron: A Highly-Effective Feature-Domain Neural Trojan Using Invisible and Polymorphic Watermarks},
  booktitle = {Proceedings of the 2022 ACM SIGSAC Conference on Computer and Communications Security},
  series    = {CCS '22},
  year      = {2022},
  isbn      = {978-1-4503-9450-5},
  location  = {Los Angeles, CA, USA},
  pages     = {2129--2143},
  numpages  = {15},
  url       = {https://dl.acm.org/doi/10.1145/3548606.3560678},
  doi       = {10.1145/3548606.3560678},
  publisher = {ACM},
  address   = {New York, NY, USA}
}

@inproceedings{he2016deep,
  title={Deep residual learning for image recognition},
  author={He, Kaiming and Zhang, Xiangyu and Ren, Shaoqing and Sun, Jian},
  booktitle={Proceedings of the IEEE conference on computer vision and pattern recognition},
  pages={770--778},
  year={2016}
}

@inproceedings{dosovitskiy2021image,
  title={An image is worth 16x16 words: Transformers for image recognition at scale},
  author={Dosovitskiy, Alexey and Beyer, Lucas and Kolesnikov, Alexander and Weissenborn, Dirk and Zhai, Xiaohua and Unterthiner, Thomas and Dehghani, Mostafa and Minderer, Matthias and Heigold, Georg and Gelly, Sylvain and others},
  booktitle={International Conference on Learning Representations},
  year={2021}
}

@inproceedings{molchanov2016pruning,
  title={Pruning convolutional neural networks for resource efficient inference},
  author={Molchanov, Pavlo and Tyree, Stephen and Karras, Tero and Aila, Timo and Kautz, Jan},
  booktitle={International Conference on Learning Representations},
  year={2017}
}

@inproceedings{moosavi2017universal,
  title={Universal adversarial perturbations},
  author={Moosavi-Dezfooli, Seyed-Mohsen and Fawzi, Alhussein and Fawzi, Omar and Frossard, Pascal},
  booktitle={Proceedings of the IEEE conference on computer vision and pattern recognition},
  pages={1765--1773},
  year={2017}
}

@techreport{openai2024gpt4,
  title={GPT-4 Technical Report},
  author={{OpenAI}},
  institution={OpenAI},
  year={2024},
  url={https://arxiv.org/abs/2303.08774},
  note={arXiv:2303.08774v6}
}

@article{touvron2023llama,
  title={LLaMA: Open and Efficient Foundation Language Models},
  author={Touvron, Hugo and Lavril, Thibaut and Izacard, Gautier and Martinet, Xavier and Lachaux, Marie-Anne and Lacroix, Timoth{\'e}e and Rozi{\`e}re, Baptiste and Goyal, Naman and Hambro, Eric and Azhar, Faisal and Rodriguez, Aurelien and Joulin, Armand and Grave, Edouard and Lample, Guillaume},
  journal={arXiv preprint arXiv:2302.13971},
  year={2023},
  url={https://arxiv.org/abs/2302.13971}
}

@inproceedings{adi2018turning,
  title={Turning your weakness into a strength: Watermarking deep neural networks by backdooring},
  author={Adi, Yossi and Baum, Carsten and Cisse, Moustapha and Pinkas, Benny and Keshet, Joseph},
  booktitle={27th USENIX Security Symposium (USENIX Security 18)},
  pages={1615--1631},
  year={2018}
}

@inproceedings{zhang2018protecting,
  title={Protecting intellectual property of deep neural networks with watermarking},
  author={Zhang, Jialong and Gu, Zhongshu and Jang, Jiyong and Wu, Hui and Stoecklin, Marc Ph and Huang, Heqing and Molloy, Ian},
  booktitle={Proceedings of the 2018 on Asia Conference on Computer and Communications Security},
  pages={159--172},
  year={2018}
}

@inproceedings{uchida2017embedding,
  title={Embedding watermarks into deep neural networks},
  author={Uchida, Yuki and Nagai, Yuki and Sakazawa, Shigeyuki and Satoh, Shin'ichi},
  booktitle={Proceedings of the 2017 ACM on International Conference on Multimedia Retrieval},
  pages={269--277},
  year={2017}
}

@inproceedings{chen2021model,
  title={Model Assertion: A Defense Against Model Theft via Authorized Model Encryption},
  author={Chen, Huili and Fu, Cheng and Zhao, Jishen and Koushanfar, Farinaz},
  booktitle={Proceedings of the IEEE/CVF International Conference on Computer Vision (ICCV)},
  pages={15380--15389},
  year={2021}
}

@inproceedings{rouhani2019deepsigns,
  title={DeepSigns: An End-to-End Watermarking Framework for Ownership Protection of Deep Neural Networks},
  author={Rouhani, Bita Darvish and Chen, Huili and Koushanfar, Farinaz},
  booktitle={Proceedings of the Twenty-Fourth International Conference on Architectural Support for Programming Languages and Operating Systems (ASPLOS)},
  pages={485--497},
  year={2019},
  organization={ACM},
  doi={10.1145/3297858.3304051}
}

@inproceedings{li2022untargeted,
  title={Untargeted Backdoor Watermark: Towards Harmless and Stealthy Dataset Copyright Protection},
  author={Li, Yiming and Bai, Yang and Jiang, Yong and Yang, Yong and Xia, Shu-Tao and Li, Bo},
  booktitle={Advances in Neural Information Processing Systems (NeurIPS)},
  volume={35},
  pages={13862--13875},
  year={2022},
  note={Oral, Top 2\%}
}

@inproceedings{lukas2022sok,
  title={SoK: How Robust is Image Classification Deep Neural Network Watermarking?},
  author={Lukas, Nils and Jiang, Edward and Li, Xinda and Kerschbaum, Florian},
  booktitle={2022 IEEE Symposium on Security and Privacy (SP)},
  pages={787--804},
  year={2022},
  organization={IEEE}
}

@inproceedings{guo2023domain,
  title={Domain Watermark: Effective and Harmless Dataset Copyright Protection is Closed at Hand},
  author={Guo, Junfeng and Li, Yiming and Wang, Lixu and Xia, Shu-Tao and Huang, Heng and Liu, Cong and Li, Bo},
  booktitle={Advances in Neural Information Processing Systems (NeurIPS)},
  year={2023}
}

@inproceedings{tramer2019slalom,
  title={Slalom: Fast, Verifiable and Private Execution of Neural Networks in Trusted Hardware},
  author={Tramer, Florian and Boneh, Dan},
  booktitle={International Conference on Learning Representations (ICLR)},
  year={2019}
}

@article{feng2024survey,
  title={Survey of Research on Confidential Computing},
  author={Feng, Wei and others},
  journal={IET Communications},
  volume={18},
  number={8},
  pages={465--486},
  year={2024},
  publisher={Wiley},
  doi={10.1049/cmu2.12759}
}

@inproceedings{natarajan2023chexmix,
  title={CHEX-MIX: Combining Homomorphic Encryption with Trusted Execution Environments for Two-party Oblivious Inference in the Cloud},
  author={Natarajan, Deepika and Loveless, Andrew and Dai, Wei and Dreslinski, Ronald},
  booktitle={8th IEEE European Symposium on Security and Privacy (EuroS\&P)},
  year={2023},
  pages={457--477}
}

@article{xue2023activeguard,
  title={ActiveGuard: An Active Intellectual Property Protection Technique for Deep Neural Networks by Leveraging Adversarial Examples as Users' Fingerprints},
  author={Xue, Mingfu and Sun, Shichang and He, Can and Gu, Dujuan and Zhang, Yushu and Wang, Jian and Liu, Weiqiang},
  journal={IET Computers \& Digital Techniques},
  volume={17},
  number={3-4},
  pages={111--126},
  year={2023},
  publisher={Wiley},
  doi={10.1049/cdt2.12056}
}

@article{xue2024ssat,
  title={SSAT: Active Authorization Control and User's Fingerprint Tracking Framework for DNN IP Protection},
  author={Xue, Mingfu and Wu, Yinghao and Zhang, Leo Yu and Gu, Dujuan and Zhang, Yushu and Liu, Weiqiang},
  journal={ACM Transactions on Multimedia Computing, Communications and Applications},
  volume={20},
  number={10},
  year={2024},
  publisher={ACM},
  doi={10.1145/3679202}
}

@inproceedings{mopuri2018generalizable,
  title={Generalizable data-free objective for crafting universal adversarial perturbations},
  author={Mopuri, Konda Reddy and Ganeshan, Aditya and Babu, R Venkatesh},
  booktitle={IEEE Transactions on Pattern Analysis and Machine Intelligence (TPAMI)},
  volume={41},
  number={10},
  pages={2452--2465},
  year={2019}
}

@inproceedings{li2016pruning,
  title={Pruning filters for efficient convnets},
  author={Li, Hao and Kadav, Asim and Durdanovic, Igor and Samet, Hanan and Graf, Hans Peter},
  booktitle={International Conference on Learning Representations (ICLR)},
  year={2017}
}

@inproceedings{molchanov2019importance,
  title={Importance estimation for neural network pruning},
  author={Molchanov, Pavlo and Mallya, Arun and Tyree, Stephen and Frosio, Iuri and Kautz, Jan},
  booktitle={Proceedings of the IEEE/CVF Conference on Computer Vision and Pattern Recognition (CVPR)},
  pages={11264--11272},
  year={2019}
}

@inproceedings{han2015learning,
  title={Learning both weights and connections for efficient neural networks},
  author={Han, Song and Pool, Jeff and Tran, John and Dally, William J},
  booktitle={Advances in Neural Information Processing Systems (NeurIPS)},
  pages={1135--1143},
  year={2015}
}

@article{chen2017targeted,
  title={Targeted Backdoor Attacks on Deep Learning Systems Using Data Poisoning},
  author={Chen, Xinyun and Liu, Chang and Li, Bo and Lu, Kimberly and Song, Dawn},
  journal={arXiv preprint arXiv:1712.05526},
  year={2017}
}

@article{gu2017badnets,
  title={BadNets: Identifying Vulnerabilities in the Machine Learning Model Supply Chain},
  author={Gu, Tianyu and Dolan-Gavitt, Brendan and Garg, Siddharth},
  journal={arXiv preprint arXiv:1708.06733},
  year={2017}
}

@inproceedings{krizhevsky2012imagenet,
  title={Imagenet classification with deep convolutional neural networks},
  author={Krizhevsky, Alex and Sutskever, Ilya and Hinton, Geoffrey E},
  booktitle={Advances in Neural Information Processing Systems},
  volume={25},
  year={2012}
}

@inproceedings{huang2017densely,
  title={Densely Connected Convolutional Networks},
  author={Huang, Gao and Liu, Zhuang and Van Der Maaten, Laurens and Weinberger, Kilian Q},
  booktitle={Proceedings of the IEEE Conference on Computer Vision and Pattern Recognition (CVPR)},
  pages={4700--4708},
  year={2017}
}

@techreport{krizhevsky2009learning,
  title={Learning multiple layers of features from tiny images},
  author={Krizhevsky, Alex and Hinton, Geoffrey and others},
  year={2009},
  institution={University of Toronto}
}

@article{lecun1998gradient,
  title={Gradient-based learning applied to document recognition},
  author={LeCun, Yann and Bottou, L{\'e}on and Bengio, Yoshua and Haffner, Patrick},
  journal={Proceedings of the IEEE},
  volume={86},
  number={11},
  pages={2278--2324},
  year={1998},
  publisher={IEEE}
}

@inproceedings{netzer2011reading,
  title={Reading digits in natural images with unsupervised feature learning},
  author={Netzer, Yuval and Wang, Tao and Coates, Adam and Bissacco, Alessandro and Wu, Bo and Ng, Andrew Y},
  booktitle={NIPS Workshop on Deep Learning and Unsupervised Feature Learning},
  year={2011}
}

@article{xue2021advparams,
  title={AdvParams: An Active DNN Intellectual Property Protection Technique via Adversarial Perturbation Based Parameter Encryption},
  author={Xue, Mingfu and Wu, Zhushou and Wang, Jian and Zhang, Yushu and Liu, Weiqiang},
  journal={arXiv preprint arXiv:2105.13697},
  year={2021}
}

@inproceedings{alam2020deeplock,
  title={Deep-Lock: Secure Authorization for Deep Neural Networks},
  author={Alam, Manaar and Saha, Sayandeep and Mukhopadhyay, Debdeep and Kundu, Sandip},
  booktitle={2020 IEEE 38th VLSI Test Symposium (VTS)},
  pages={1--6},
  year={2020},
  organization={IEEE}
}

@article{alam2022nnlock,
  title={NN-Lock: A Lightweight Authorization to Prevent IP Threats of Deep Learning Models},
  author={Alam, Manaar and Saha, Sayandeep and Mukhopadhyay, Debdeep and Kundu, Sandip},
  journal={ACM Journal on Emerging Technologies in Computing Systems},
  volume={18},
  number={2},
  pages={1--27},
  year={2022},
  publisher={ACM}
}

@inproceedings{chakraborty2020hpnn,
  title={Hardware-Assisted Intellectual Property Protection of Deep Learning Models},
  author={Chakraborty, Abhishek and Mondal, Ankit and Srivastava, Ankur},
  booktitle={2020 57th ACM/IEEE Design Automation Conference (DAC)},
  pages={1--6},
  year={2020},
  organization={IEEE}
}

@inproceedings{gong2024modellock,
  title={ModelLock: Locking Your Model With a Spell},
  author={Gong, Yifan and Chen, Dongliang and Niu, Weizhan and Cheng, Shuai and Pan, Xiaohang and Nie, Qingyuan and Xiao, Yanjiao and Zhang, Linghe and Zheng, Haoyu},
  booktitle={Proceedings of the 32nd ACM International Conference on Multimedia},
  pages={6595--6604},
  year={2024}
}

@inproceedings{liu2023enhancing,
  title={Enhancing Generalization of Universal Adversarial Perturbation through Gradient Aggregation},
  author={Liu, Xuannan and Zhong, Yaoyao and Zhang, Yuhang and Qin, Lixiong and Deng, Weihong},
  booktitle={Proceedings of the IEEE/CVF International Conference on Computer Vision (ICCV)},
  pages={4428--4437},
  year={2023}
}

@inproceedings{zhang2025improving,
  title={Improving Generalization of Universal Adversarial Perturbation via Dynamic Maximin Optimization},
  author={Zhang, Yechao and Xu, Yingzhe and Shi, Junyu and Zhang, Leo Yu and Hu, Shengshan and Li, Minghui and Zhang, Yanjun},
  booktitle={Proceedings of the AAAI Conference on Artificial Intelligence},
  volume={39},
  pages={--},
  year={2025}
}

@article{xu2022robust,
  title={Robust Universal Adversarial Perturbations},
  author={Xu, Changming and Singh, Gagandeep},
  journal={arXiv preprint arXiv:2206.10858},
  year={2022}
}

@article{zhong2024sparse,
  title={Sparse-{PGD}: A Unified Framework for Sparse Adversarial Perturbations Generation},
  author={Zhong, Xuyang and Liu, Chen},
  journal={arXiv preprint arXiv:2405.05075},
  year={2024}
}

@inproceedings{cao2021ipguard,
  title={{IPGuard}: Protecting Intellectual Property of Deep Neural Networks via Fingerprinting the Classification Boundary},
  author={Cao, Xiaoyu and Jia, Jinyuan and Gong, Neil Zhenqiang},
  booktitle={Proceedings of the 2021 ACM Asia Conference on Computer and Communications Security (AsiaCCS)},
  pages={14--25},
  year={2021},
  doi={10.1145/3433210.3437526}
}

@article{lemerrer2020adversarial,
  title={Adversarial Frontier Stitching for Remote Neural Network Watermarking},
  author={Le Merrer, Erwan and P{\'e}rez, Patrick and Tr{\'e}dan, Gilles},
  journal={Neural Computing and Applications},
  volume={32},
  number={13},
  pages={9233--9244},
  year={2020},
  publisher={Springer},
  doi={10.1007/s00521-019-04434-z}
}

@article{fan2022deepipr,
  title={DeepIPR: Deep Neural Network Ownership Verification With Passports},
  author={Fan, Lixin and Ng, Kam Woh and Chan, Chee Seng},
  journal={IEEE Transactions on Pattern Analysis and Machine Intelligence},
  volume={44},
  number={10},
  pages={6122--6139},
  year={2022},
  publisher={IEEE}
}

@inproceedings{tramer2016stealing,
  title={Stealing Machine Learning Models via Prediction {APIs}},
  author={Tram{\`e}r, Florian and Zhang, Fan and Juels, Ari and Reiter, Michael K and Ristenpart, Thomas},
  booktitle={25th USENIX Security Symposium (USENIX Security 16)},
  pages={601--618},
  year={2016},
  publisher={USENIX Association}
}

@inproceedings{jagielski2020high,
  title={High accuracy and high fidelity extraction of neural networks},
  author={Jagielski, Matthew and Carlini, Nicholas and Berthelot, David and Kurakin, Alex and Papernot, Nicolas},
  booktitle={29th USENIX Security Symposium (USENIX Security 20)},
  pages={1345--1362},
  year={2020},
  publisher={USENIX Association}
}

@inproceedings{fredrikson2015model,
  title={Model inversion attacks that exploit confidence information and basic countermeasures},
  author={Fredrikson, Matt and Jha, Somesh and Ristenpart, Thomas},
  booktitle={Proceedings of the 22nd ACM SIGSAC Conference on Computer and Communications Security (CCS)},
  pages={1322--1333},
  year={2015},
  publisher={ACM}
}

@inproceedings{orekondy19knockoff,
    TITLE = {Knockoff Nets: Stealing Functionality of Black-Box Models},
    AUTHOR = {Orekondy, Tribhuvanesh and Schiele, Bernt and Fritz, Mario},
    YEAR = {2019},
    BOOKTITLE = {CVPR},
}

@inproceedings{zhang2020secret,
  author    = {Yuheng Zhang and Ruoxi Jia and Hengzhi Pei and Wenxiao Wang 
               and Bo Li and Dawn Song},
  title     = {The Secret Revealer: Generative Model-Inversion Attacks 
               Against Machine Learning Models},
  booktitle = {Proceedings of the IEEE/CVF Conference on Computer Vision 
               and Pattern Recognition (CVPR)},
  year      = {2020},
  pages     = {253--261}
}

@inproceedings{orekondy2020prediction,
   title={Prediction Poisoning: Towards Defenses Against {DNN} Model Stealing Attacks},
   author={Orekondy, Tribhuvanesh and Schiele, Bernt and Fritz, Mario},
   booktitle={International Conference on Learning Representations (ICLR)},
   year={2020}
 }

@article{zhang2023cip,
   title={Categorical Inference Poisoning: Verifiable Defense Against Black-Box {DNN} Model Stealing Without Constraining Surrogate Data and Query Times},
   author={Zhang, Haitian and Hua, Guang and Wang, Xinya and Jiang, Hao and Yang, Wen},
   journal={IEEE Transactions on Information Forensics and Security},
   volume={18},
   pages={1473--1486},
   year={2023}
 }

@article{jiang2024amao,
   title={{AMAO}: A Comprehensive Defense Framework Against Model Extraction Attacks},
   author={Jiang, Mingyi and others},
   journal={IEEE Transactions on Dependable and Secure Computing},
   volume={21},
   number={2},
   year={2024}
 }

@article{chakraborty2022dynamarks,
   title={{DynaMarks}: Defending Against Deep Learning Model Extraction Using Dynamic Watermarking},
   author={Chakraborty, Abhishek and others},
   journal={arXiv preprint arXiv:2207.13321},
   year={2022}
 }

\end{document}